\documentclass[aps,onecolumn,12pt,floatfix,altaffilletter,preprintnumbers,nofootinbib,%
               tightenlines,showpacs,showkeys,superscriptaddress]{revtex4-1} 

\usepackage[colorlinks=true,citecolor=blue,linkcolor=blue]{hyperref}
\usepackage[utf8]{inputenc}
\usepackage[normalem]{ulem}
\usepackage{amsmath,amssymb}
\usepackage{epsfig}
\usepackage{graphicx}               
\usepackage{url}
\usepackage{color}
\usepackage{multirow}
\usepackage{placeins}
\usepackage[dvipsnames]{xcolor}
\usepackage{epstopdf}
\usepackage{hhline}
\usepackage{multirow}

\usepackage{tikz}
\usetikzlibrary{trees}
\usetikzlibrary{decorations.pathmorphing}
\usetikzlibrary{decorations.markings}

\definecolor{jblue}  {RGB}{20,50,100}
\definecolor{npurple}  {RGB} {153, 51, 204}
\definecolor{wred}   {RGB}{217,0,56}
\definecolor{white}   {RGB}{255,255,255}

\definecolor{korange}   {RGB}{235, 80,  43}
\definecolor{korange2}   {RGB}{245, 100,  63}
\definecolor{kyelloworange}   {RGB}{255, 210,  110}
\definecolor{kyelloworange2}   {RGB}{240, 170,  90}
\definecolor{kred}   {RGB}{204,  102, 153}
\definecolor{kpurple}   {RGB}{153,  61, 190}
\definecolor{kpurplelight}   {RGB}{213,  161, 230}

\usepackage{cleveref}
\usepackage{subfigure} 
\usepackage{lipsum}
\definecolor{red}{rgb}{1.0, 0, 0}
 \usepackage{gensymb}
\allowdisplaybreaks

\bibpunct{[}{]}{,}{n}{}{,}

\newcommand{\ev}[1]{\ensuremath{\left\langle #1 %
                     \right\rangle}} 

\newcommand{\parenbar}[1]{\overset{
            \raisebox{-0.15em}{\scalebox{.4}{\textbf{(}}}
            \raisebox{-0.3em}{{\hspace{.03em}--\hspace{.05em}}}
            \raisebox{-0.15em}{\scalebox{.4}{\textbf{)}}}} {#1}}

\newcommand{\iso}[2]{{\ensuremath{{}^{#2}}\ensuremath{\rm #1}}}

\newcommand{\eV}{\,\mathrm{eV}}

\begin{document}

\preprint{MITP/18-023, FERMILAB-PUB-18-086-T, IFT-UAM/CSIC-18-033}

\title{Updated global analysis of neutrino oscillations \\
       in the presence of eV-scale sterile neutrinos}

\author{Mona~Dentler}        \email{modentle@uni-mainz.de}
\affiliation{PRISMA Cluster of Excellence and
             Mainz Institute for Theoretical Physics,
             Johannes Gutenberg-Universit\"{a}t Mainz, 55099 Mainz, Germany}

\author{\'Alvaro~Hern\'andez-Cabezudo} \email{alvaro.cabezudo@kit.edu} 
\affiliation{Institut f\"ur Kernphysik,
             Karlsruher Institut f\"ur Technologie (KIT),
             76021 Karlsruhe, Germany}

\author{Joachim~Kopp}        \email{jkopp@uni-mainz.de}
\affiliation{PRISMA Cluster of Excellence and
             Mainz Institute for Theoretical Physics,
             Johannes Gutenberg-Universit\"{a}t Mainz, 55099 Mainz, Germany}
\affiliation{Theoretical Physics Department, CERN, 1211 Geneva, Switzerland}

\author{Pedro~Machado}       \email{pmachado@fnal.gov}
\affiliation{Theoretical Physics Department,
             Fermi National Accelerator Laboratory,
             Batavia, IL, 60510, USA}

\author{Michele~Maltoni}     \email{michele.maltoni@csic.es}
\author{Ivan~Martinez-Soler} \email{ivanj.martinez@estudiante.uam.es}
\affiliation{Instituto de F\'isica Te\'orica UAM/CSIC,
             Calle de Nicol\'as Cabrera 13-15, 28049 Madrid, Spain}

\author{Thomas~Schwetz}      \email{schwetz@kit.edu}
\affiliation{Institut f\"ur Kernphysik,
             Karlsruher Institut f\"ur Technologie (KIT),
             76021 Karlsruhe, Germany}

\begin{abstract}
  \vskip 1cm We discuss the possibility to explain the anomalies in
  short-baseline neutrino oscillation experiments in terms of sterile
  neutrinos.  We work in a $3+1$ framework and pay special attention to recent
  new data from reactor experiments, IceCube and MINOS+. We find that results
  from the DANSS and NEOS reactor experiments support the sterile neutrino
  explanation of the reactor anomaly, based on an analysis that relies solely on
  the relative comparison of measured reactor spectra.  Global data from the
  $\nu_e$ disappearance channel favour sterile neutrino oscillations at the
  $3\sigma$ level with $\Delta m^2_{41} \approx 1.3$~eV$^2$ and $|U_{e4}|
  \approx 0.1$, even without any assumptions on predicted reactor fluxes.  In
  contrast, the anomalies in the $\nu_e$ appearance channel (dominated by LSND)
  are in strong tension with improved bounds on $\nu_\mu$ disappearance, mostly
  driven by MINOS+ and IceCube.  Under the sterile neutrino oscillation
  hypothesis, the $p$-value for those data sets being consistent is less than
  $2.6\times 10^{-6}$. Therefore, an explanation of the LSND anomaly in terms
  of sterile neutrino oscillations in the $3+1$ scenario is excluded at the
  $4.7\sigma$ level. This result is robust with respect to variations in the
  analysis and used data, in particular it depends neither on the theoretically
  predicted reactor neutrino fluxes, nor on constraints from
  any single experiment.  Irrespective of the anomalies, we provide updated
  constraints on the allowed mixing strengths $|U_{\alpha 4}|$ ($\alpha =
  e,\mu,\tau$) of active neutrinos with a fourth neutrino mass state in the eV
  range.
\end{abstract}

\maketitle

\section{Introduction}
\label{sec:intro}

For almost two decades, the possible existence of light sterile neutrinos---new
species of neutral fermions participating in neutrino oscillation---has
intrigued the neutrino physics community.  The excitement is fuelled in
particular by a number of unexpected experimental results: an unexplained
excess of electron anti-neutrinos ($\bar\nu_e$) in a muon anti-neutrino
($\bar\nu_\mu$) beam observed at a baseline of $\sim 30$\,m from the source in
the LSND experiment~\cite{Aguilar:2001ty}; a similar excess found by the
MiniBooNE collaboration at higher energies and correspondingly larger
baseline~\cite{Aguilar-Arevalo:2013pmq}; the disagreement between theoretically
predicted $\bar\nu_e$ fluxes from nuclear reactors and observations
\cite{Mueller:2011nm, Huber:2011wv}, known as the reactor anti-neutrino anomaly
\cite{Mention:2011rk} (see also \cite{Hayes:2013wra, Fang:2015cma,
Hayes:2016qnu}); and a similar disagreement between expectations and
observations in experiments using intense radioactive sources
\cite{Acero:2007su, Giunti:2010zu}.  

These anomalies need to be contrasted with a large set of null results in the
$\nu_\mu \to \nu_\mu$, $\nu_e \to \nu_e$, and $\nu_\mu \to \nu_e$ oscillation
channels as well as the corresponding anti-neutrino channels.
The observation of all of these channels overconstrains sterile neutrino
models, therefore global fits of such models exhibit pronounced
tension, even though different data sets on each individual oscillation channel
are consistent, for recent analyses see e.g.~\cite{Kopp:2011qd, Conrad:2012qt,
Archidiacono:2013xxa, Kopp:2013vaa, Mirizzi:2013kva, Giunti:2013aea,
Gariazzo:2013gua, Collin:2016rao, Gariazzo:2017fdh, Giunti:2017yid,
Dentler:2017tkw}.

In this work, we update our previous analyses from
refs.~\cite{Kopp:2011qd, Kopp:2013vaa, Dentler:2017tkw} to incorporate new
experimental results.  These are in particular the following:
\begin{enumerate}
  \item New constraints on $\bar\nu_e$ disappearance into sterile
    neutrinos from the reactor neutrino experiments Daya
    Bay~\cite{An:2016luf}, NEOS \cite{Ko:2016owz}, and DANSS
    \cite{Alekseev:2016llm, danss-moriond17,danss-solvay17}.  Unlike
    the results from previous short-baseline reactor experiments that
    have led to the reactor anti-neutrino anomaly, these new analyses
    are based on a comparison of measured spectra at different
    baselines rather than a comparison of data to theoretically
    predicted spectra. The new results are therefore insensitive to
    possible mismodelling of the $\bar\nu_e$ emission from nuclear
    reactors.  In particular, they are insensitive to an observed, but
    so far unexplained, bump at neutrino energies $\sim 5$~MeV
    \cite{Seo:2016uom, Abe:2014bwa, An:2016srz}\footnote{See
      refs.~\cite{Dwyer:2014eka, Hayes:2015yka, Novella:2015eaw,
      Giunti:2016elf, Huber:2016xis, Mohanty:2017iyh} for a discussion of
      possible nuclear physics or experimental origins of this bump, and
      ref.~\cite{Berryman:2018jxt} for speculations about a possible
      new physics explanation.}.  Spectral distortions in
    the recent data from DANSS and NEOS lead to a hint in favour of
    sterile neutrinos at the $3\sigma$ level, which supports the
    previous reactor anomaly independent of flux predictions.

  \item Daya Bay measurements of the individual neutrino fluxes from
    different fissible isotopes~\cite{An:2017osx}.  By combining the
    time evolution of the observed reactor anti-neutrino spectra with
    the known evolution of the reactor fuel composition, the Daya Bay
    collaboration was able to determine independently the neutrino
    fluxes from the two most important fissible isotopes in a nuclear
    reactor, \iso{U}{235} and \iso{Pu}{239}.  Their analysis suggests
    that the discrepancy between predicted and observed fluxes stems
    mainly from \iso{U}{235}, while the neutrino flux from
    \iso{Pu}{239} appears consistent with predictions.  (The other
    potentially relevant isotopes \iso{U}{238} and \iso{Pu}{241} are
    subdominant in Daya Bay.) In contrast, oscillations into sterile
    neutrinos would lead to equal flux deficits in all isotopes.
    Implications of these results for sterile neutrino models have
    been discussed previously in refs.~\cite{Giunti:2017yid,
      Dentler:2017tkw}. In our previous paper~\cite{Dentler:2017tkw}
    we have shown that both hypotheses (free flux normalizations
    versus sterile neutrino oscillations) give acceptable fits to Daya
    Bay data, and that the preference in favour of flux rescaling
    decreases once Daya Bay is combined with the global reactor
    data. We will update those results in \cref{sec:reactor} below.
    Finally, it has been demonstrated recently that the theoretical
    predictions for the time-dependence of reactor anti-neutrino
    fluxes on which the Daya Bay analysis is based may need to be
    refined \cite{Huber:2015ouo, Huber:nuplatformweek}. In particular,
    the present analysis accounts neither for the time-dependent
    equilibration of decay chains nor for the possibility of neutron
    capture on fission products, which would lead to a non-linear
    dependence of anti-neutrino fluxes on the neutron flux in the
    reactor~\cite{Huber:2015ouo}. Taking these effects into account,
    Daya Bay's preference for the flux misprediction hypothesis is
    estimated to drop to well below
    $2\sigma$~\cite{Huber:nuplatformweek}.

  \item Final results from OPERA~\cite{Agafonova:2013xsk} and
    ICARUS~\cite{Antonello:2012fu, Farese:2014}. 
    Both experiments constrain sterile neutrinos mixing with electron and
    muon neutrinos by searching for anomalous $\nu_\mu \to \nu_e$ appearance in
    the CNGS beam.

  \item Searches for sterile neutrinos in MINOS/MINOS+ \cite{Adamson:2017uda}
    and in NO$\nu$A \cite{Adamson:2017zcg}. The first analysis combines charged current
    $\nu_\mu$ disappearance data and neutral current data from
    the MINOS experiment and from the MINOS+ setup operating the same detector
    in a higher energy beam. The second analysis is based on neutral current data
    from NO$\nu$A. Especially the MINOS/MINOS+ analysis places stringent
    bounds on sterile neutrino mixing with $\nu_\mu$ over a wide range of masses.

  \item New solar neutrino data, including the 2055-day energy and
    day/night asymmetry spectrum from Super-Kamiokande
    phase~4~\cite{sksol:nakano2016} and the measurement of neutrinos
    from the proton-proton ($pp$) fusion chain in the Sun recently
    presented by Borexino~\cite{Bellini:2014uqa}. In addition, the
    results of all solar experiments have been updated to match the
    new solar neutrino fluxes predicted by the GS98 version of the
    Standard Solar Model presented in ref.~\cite{Vinyoles:2016djt}.

  \item Improved atmospheric neutrino data from Super-Kamiokande
    (including 1775 days of phase~4 data) from ref.~\cite{wendell:2014dka}, as
    well as the complete set of DeepCore 3-year data presented in
    ref.~\cite{Aartsen:2014yll} and publicly released in
    ref.~\cite{deepcore:2016}. The calculations of atmospheric neutrino event
    rates for both detectors are based on the atmospheric neutrino flux
    calculations described in ref.~\cite{Honda:2015fha}.
    
  \item First sterile neutrino limits from IceCube, based on one year
    of data~\cite{TheIceCube:2016oqi, Jones:2015, Arguelles:2015}.
    This novel analysis exploits the fact that active-to-sterile
    oscillations of atmospheric neutrinos inside the Earth may be
    enhanced by a Mikheyev-Smirnov-Wolfenstein (MSW)
    resonance~\cite{Wolfenstein:1977ue, Mikheev:1986gs}. The resonance
    affects the anti-neutrino sector, and for sterile neutrino masses
    around 1~eV occurs at energies of order 1~TeV, an energy well
    above IceCube's detection threshold, but still low enough to
    benefit from a substantial flux~\cite{Nunokawa:2003ep,
      Choubey:2007ji}. Consequently, IceCube is able to set strong
    limits on sterile neutrino mixing with $\nu_\mu$.
\end{enumerate}

We will begin in \cref{sec:osc} by reviewing the formalism of neutrino
oscillations in the presence of sterile neutrinos. Along the way, we
will also fix our notation, such as our parameterization of the
leptonic mixing matrix. In
\cref{sec:nu-e,sec:nu-mu-to-nu-e,sec:nu-mu}, we will then discuss the
status of the global data sets in the $\nu_e \to \nu_e$, $\nu_\mu \to
\nu_e$, and $\nu_\mu \to \nu_\mu$ channels (and the corresponding
anti-neutrino channels) in turn. In particular, \cref{sec:nu-e}
discusses the recent hints from reactor spectral data and
\cref{sec:nu-mu-to-nu-e} reviews the anomalies in the appearance
channel.  In \cref{sec:nu-mu,sec:U-tau-4}, we present updated
constraints on the mixing of a sterile neutrino with the $\nu_\mu$ and
$\nu_\tau$ flavour from global data, respectively. We will finally
combine all oscillation channels in \cref{sec:gof} into a global fit.
We will determine the goodness of fit at the global best fit point and
quantify the tension between appearance and disappearance data. We
will summarize our results and conclude in
\cref{sec:discussion}. Supplementary material can be found in the
appendices.

\section{Neutrino Oscillations in the Presence of Sterile Neutrinos}
\label{sec:osc}

The topic of this paper are scenarios in which the standard three-flavor
framework for neutrino oscillations is augmented by adding one sterile
neutrinos $\nu_s$. We will refer to such scenarios as ``$3+1$ models''.
We will comment on scenarios with more than one sterile neutrino
in \cref{sec:discussion}.

The oscillation probability for $\nu_\alpha \to \nu_\beta$
transitions in vacuum ($\alpha, \beta = e, \mu, \tau, s$) is given by
\begin{align}
  P_{\alpha\beta} = \sum_{j,k=1}^4 U_{\alpha j}^* U_{\beta j}
    U_{\alpha k} U_{\beta k}^* \exp\bigg[-i \frac{\Delta m_{jk}^2 L}{2 E} \bigg] \,.
  \label{eq:P}
\end{align}
Here, $L$ is the baseline, $E$ is the neutrino energy, $U_{\alpha j}$
are the elements of the leptonic mixing matrix (which is $4 \times 4$
in a $3+1$ model), and $\Delta m_{jk}^2 \equiv m_j^2 - m_k^2$ are the
mass squared differences, with $m_j$ the neutrino mass eigenvalues.
We will assume $m_{1,2,3} \ll 1\,\text{eV}$, but allow $m_4$ to be
larger, thus considering the case $\Delta m_{41}^2 > 0$.
For experiments in which matter effects play a significant
role, in general the evolution equation should be solved numerically.
In cases where a constant matter density is a good approximation,
$U_{\alpha j}$ and $\Delta m_{jk}^2$ in \cref{eq:P} can be replaced by
an effective mixing matrix and effective mass squared differences in
matter.  For anti-neutrino oscillations, $U$ should be replaced by
$U^*$.

The mixing matrix $U$ in vacuum can be written as a product of two-dimensional
rotation matrices. Where an explicit parameterization is required, we choose
\begin{align}
  U \equiv R_{34}(\theta_{34}) \, R_{24}(\theta_{24}, \delta_{24}) \, 
      R_{14}(\theta_{14}) \, R_{23}(\theta_{23}) \,
      R_{13}(\theta_{13}, \delta_{13}) \, R_{12}(\theta_{12}, \delta_{12}) \,,
  \label{eq:U-parameterization}
\end{align}
where $R_{ij}(\theta_{ij})$ denotes a real rotation matrix in the $(ij)$-plane
with rotation angle $\theta_{ij}$, and $R_{ij}(\theta_{ij}, \delta_{ij})$
includes in addition a complex phase $\delta_{ij}$.  In most cases, however,
we will present our results in terms of the parameterization-independent
matrix elements $U_{\alpha\beta}$.

For the following discussion the so-called short-baseline limit of
\cref{eq:P} will be useful. This limit refers to the situation where
$\Delta m_{21}^2 L / 4 E \ll 1$, $\Delta m_{31}^2 L / 4 E \ll 1$,
so that standard three-flavor oscillations have not had time to develop yet. In
this case, \cref{eq:P} generically simplifies to
\begin{align}
  P_{\alpha\alpha}^\text{SBL}
    &= 1 - 4 |U_{\alpha 4}|^2 (1 - |U_{\alpha 4}|^2) \,
  \sin^2 \bigg(\frac{\Delta m_{41}^2 L}{4 E} \bigg) \,,
    \label{eq:P-SBL-dis}\\
  P_{\alpha\beta}^\text{SBL}
    &= 4 |U_{\alpha 4}|^2 |U_{\beta 4}|^2 \,
                             \sin^2 \bigg(\frac{\Delta m_{41}^2 L}{4 E} \bigg).
                                & (\alpha \neq \beta)
  \label{eq:P-SBL-app}
\end{align}
As we will see later, the connection between the $\nu_e\to\nu_e$,
$\nu_\mu\to\nu_\mu$, and $\nu_\mu\to\nu_e$ oscillation probabilities, inferred
from these equations, will prove to be crucial to test the compatibility
between different oscillation data sets.

An extended discussion of various other limiting cases and the
corresponding parameter dependencies (including complex phases) can be
found in ref.~\cite{Kopp:2013vaa}.

\section{$\protect\parenbar{\nu}_e$ Disappearance Data}
\label{sec:nu-e}

\begin{table}
  \centering
  \begin{ruledtabular}
  \begin{tabular}{llcp{10cm}}
    & Experiment   & References & Comments \hfill (Data points)\\
    \hline
    \multicolumn{4}{l}{{\bf Reactor experiments} \hfill (233)} \\
    & ILL          & \cite{Kwon:1981ua} \\
    & G\"osgen     & \cite{Zacek:1986cu} \\
    & Krasnoyarsk  & \cite{Vidyakin:1987ue, Vidyakin:1994ut, Kozlov:1999cs} \\
    & Rovno        & \cite{Afonin:1988gx, Kuvshinnikov:1990ry} \\
    & Bugey-3      & \cite{Declais:1994su}
                   & spectra at 3 distances with free bin-by-bin normalization\\
    & Bugey-4      & \cite{Declais:1994ma} \\
    & SRP          & \cite{Greenwood:1996pb} \\
    & NEOS         & \cite{Ko:2016owz,An:2016srz} & ratio of NEOS and Daya Bay spectra\\
    & DANSS        & \cite{danss-solvay17}
                   & ratios of spectra at two baselines
                     (updated w.r.t.~\cite{Dentler:2017tkw})\\
    & Double Chooz & \cite{Giunti:2016elf} & near detector rate \\
    & RENO         & \cite{reno-EPS17,reno-Neutrino14} & near detector rate \\
    & Daya Bay spectrum & \cite{An:2016ses} & spectral ratios EH3/EH1 and EH2/EH1 \\
    & Daya Bay flux & \cite{An:2017osx} & individual fluxes for each isotope (EH1, EH2) \\
    & KamLAND      & \cite{Gando:2010aa} & very long-baseline reactor experiment ($L \gg 1$\,km)\\
    \hline
    \multicolumn{4}{l}{{\bf Solar neutrino experiments} \hfill (325)} \\
    & Chlorine     & \cite{Cleveland:1998nv} & \\
    & GALLEX/GNO   & \cite{Kaether:2010ag} \\
    & SAGE         & \cite{Abdurashitov:2009tn} \\
    & Super-Kamiokande & \cite{Hosaka:2005um, Cravens:2008aa, Abe:2010hy, sksol:nakano2016}
                   & Phases I--IV \\
    & SNO          & \cite{Aharmim:2007nv, Aharmim:2005gt, Aharmim:2008kc}
                   & Phases 1--3 (CC and NC data) \\
    & Borexino     & \cite{Bellini:2011rx, Bellini:2008mr, Bellini:2014uqa}
                   & Phases I and II \\
    \hline
    \multicolumn{4}{l}{{\bf $\nu_e$ scattering on carbon}
      ($\nu_e + \iso{C}{12} \to \text{e}^- + \iso{N}{12}$)
      \hfill (32)} \\
    & KARMEN       & \cite{Reichenbacher:2005nc, Armbruster:1998uk, Conrad:2011ce} \\
    & LSND         & \cite{Auerbach:2001hz, Conrad:2011ce} \\
    \hline
    \multicolumn{4}{l}{{\bf Radioactive source experiments (gallium)} \hfill (4)} \\
    & GALLEX       & \cite{Hampel:1997fc, Kaether:2010ag}
                   & $\nu_e$ from \iso{Cr}{51} source \\
    & SAGE         & \cite{Abdurashitov:1998ne, Abdurashitov:2005tb}
                   & $\nu_e$ from \iso{Cr}{51} and \iso{Ar}{37} sources \\
  \end{tabular}
  \end{ruledtabular}
  \caption{Data sets included in our $\nu_e$/$\bar\nu_e$ disappearance
    analysis. The total number of data points is 594.
    More details can be found in ref.~\cite{Dentler:2017tkw}; the only
    update with respect of \cite{Dentler:2017tkw} is new data from
    DANSS \cite{danss-solvay17}.
  }
  \label{tab:nu-e-disapp}
\end{table}

In the $\nu_e$ and $\bar\nu_e$ disappearance channels, the most
important constraints on sterile neutrinos come from reactor
experiments at short baseline ($L \lesssim 1$\,km). But we include
also data from solar neutrinos, $\nu_e$ scattering on \iso{C}{12}, and
radioactive source experiments. The data is summarized in
\cref{tab:nu-e-disapp}. The following analysis is based on our earlier
publication \cite{Dentler:2017tkw} where more details can be found. In
\cref{sec:reactor} we give an update of the reactor neutrino analysis,
high-lighting the impact of the recent results from the DANSS
experiment \cite{danss-solvay17}, whereas in \cref{sec:global_nu-e} we
present the global $\protect\parenbar{\nu}_e$ disappearance analysis.

\subsection{Updated reactor analysis}
\label{sec:reactor}

The reactor analysis includes the experiments listed in
\cref{tab:nu-e-disapp}. The fit by now is dominated largely by the recent
NEOS \cite{Ko:2016owz} 
and DANSS \cite{danss-solvay17} results, as well as the latest data from Daya Bay. For
the latter we include the ratios of spectra measured in experimental
halls (EH) 3 and 1, and in experimental halls 2 and
1~\cite{An:2016ses}, as well as the measurement of the individual
neutrino fluxes from each fissible isotope~\cite{An:2017osx}.
The analysis presented here is based largely on
ref.~\cite{Dentler:2017tkw} where more details can be found. The
important difference with respect to that analysis is the recent
preliminary results from the DANSS experiment presented in December 2017
\cite{danss-solvay17}, which consists of a data sample of
approximately four times increased exposure compared to the one shown
in March~2017~\cite{danss-moriond17} used in \cite{Dentler:2017tkw}.
Another recent analysis including this latest DANSS data can be found in
ref.~\cite{Gariazzo:2018mwd}.

Regarding reactor neutrino flux predictions we consider two scenarios: (i) fixed fluxes,
where we set the uncertainties on the predicted anti-neutrino fluxes to the
values estimated in the original publications \cite{Mueller:2011nm,
Huber:2011wv}; (ii) free fluxes, where the normalizations of the neutrino
fluxes from the four main fissible isotopes \iso{U}{235}, \iso{U}{238},
\iso{Pu}{239} and \iso{Pu}{241} are allowed to float freely. (A weak constraint
$\pm 20\%$ at $1\sigma$ is included for the numerically subdominant fluxes from
\iso{U}{238} and \iso{Pu}{241} to avoid unphysical values.) Note
that we never rely on the predicted anti-neutrino \emph{spectra}, only on the
predicted rates. Even in the case of fixed fluxes, those analyses which use
spectral information are based entirely on ratios of spectra at different
baselines.

\begin{figure}
  \centering
  \hspace{-2.3cm}
  \includegraphics[width=.62\textwidth]{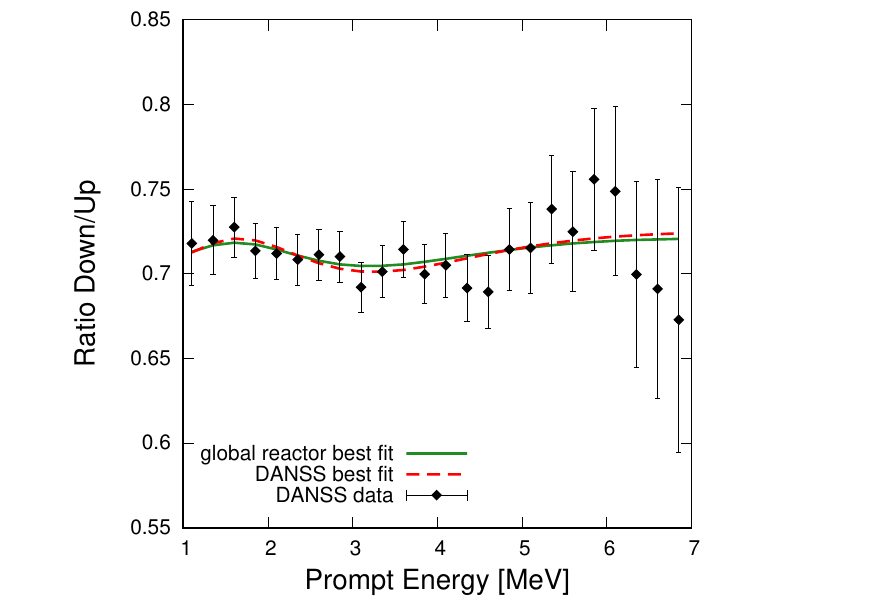}
  \hspace{-2cm}
  \includegraphics[width=.62\textwidth]{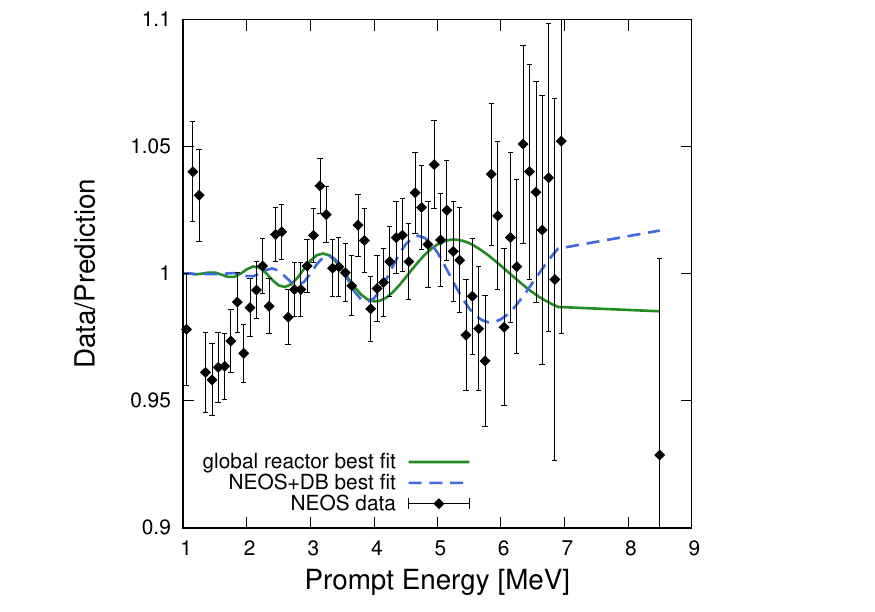}
  \hspace{-3.5cm}
  \caption{Observed spectra for the DANSS (left) and NEOS (right)
    experiments compared to the predicted spectra at the individual best fit
    points (dashed) and the best fit point from a global analysis of all
    reactor data (solid).  The left panel shows the ratio of the observed event
    rates at the two detector locations in DANSS (24 bins). The right panel
    shows the NEOS spectral data relative to the prediction extrapolated from
    the measured Day Bay spectrum (60 bins).  The best fit points are $\Delta
    m_{41}^2 = 1.32\ \text{eV}^2$, $\sin^2 \theta_{14} = 0.012$ for DANSS,
    $\Delta m_{41}^2 = 1.78\ \text{eV}^2$, $\sin^2 \theta_{14} = 0.013$ for
    NEOS + Daya Bay, and $\Delta m_{41}^2 = 1.29\ \text{eV}^2$, $\sin^2
    \theta_{14} = 0.0089$ for the fit to all reactor data, assuming a free
  normalization for the neutrino fluxes from the four main fissible isotopes.}
  \label{fig:react-spect}
\end{figure}

\begin{figure}
  \centering
  \includegraphics[width=.55\textwidth]{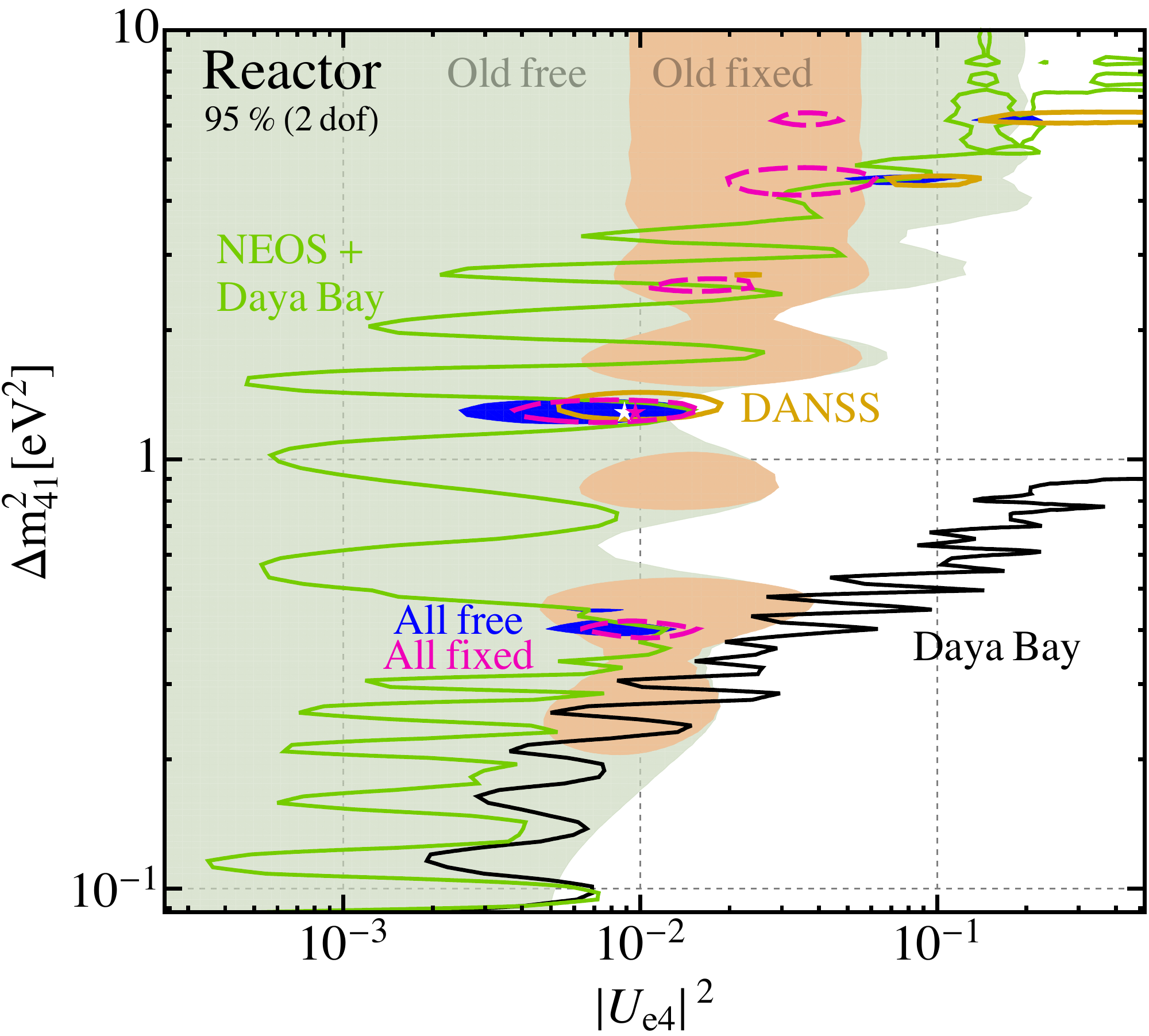}
  \caption{Allowed regions at 95\%~CL (2~dof) from reactor data. The
    solid curves correspond to Daya Bay spectral data (black), NEOS + Daya
    Bay (green), and DANSS (orange); they are independent of assumptions on fluxes
    because they are only based on spectral ratios. The light-shaded
    areas labelled ``old'' correspond to all data from
    \cref{tab:nu-e-disapp} except Daya Bay, DANSS, NEOS, and they are
    shown for the flux-free analysis making no assumptions about flux
    normalization and spectra (light orange), as well as for the flux-fixed analysis
    (light green),
    assuming reactor flux predictions and their published
    uncertainties.  The blue shaded regions correspond to all reactor
    data from \cref{tab:nu-e-disapp} for the flux-free analysis,
    whereas the dashed magenta contours indicate the global data for the
    flux fixed analysis. The white (pink) star indicates the
    best fit point $\Delta m_{41}^2 = 1.29\ \text{eV}^2$, $\sin^2
    \theta_{14} = 0.0089$  ($\Delta m_{41}^2 = 1.29\ \text{eV}^2$, $\sin^2
    \theta_{14} = 0.0096$) for free (fixed) reactor fluxes.}
  \label{fig:react-regions}
\end{figure}

\begin{table}
  \centering
  \begin{tabular}{lcc@{\quad}ccc}
    \hline\hline
    Analysis & $\Delta m^2_{41}$ [eV$^2$] & $|U^2_{e4}|$ & $\chi^2_\text{min}/\text{dof}$ & $\Delta\chi^2(\text{no-osc})$ & significance \\
    \hline
    DANSS+NEOS               & 1.3 & 0.00964 & $74.4/(84-2)$   & 13.6 & 3.3$\sigma$ \\
    all reactor (flux-free)  & 1.3 & 0.00887 & $185.8/(233-5)$ & 11.5 & 2.9$\sigma$ \\
    all reactor (flux-fixed) & 1.3 & 0.00964 & $196.0/(233-3)$ & 15.5 & 3.5$\sigma$\\
    $\parenbar\nu_e$ disap. (flux-free)
                             & 1.3 & 0.00901 & $542.9/(594-8)$ & 13.4 & 3.2$\sigma$\\
    $\parenbar\nu_e$ disap. (flux-fixed)
                             & 1.3 & 0.0102  & $552.8/(594-6)$ & 17.5 & 3.8$\sigma$\\
    \hline\hline
  \end{tabular}
  \caption{Results on $\protect\parenbar\nu_e$ disappearance
    from DANSS+NEOS, from a fit to all reactor data (both for free fluxes and
    fixed fluxes), and from a fit to the combined $\protect\parenbar\nu_e$
    disappearance data listed in \cref{tab:nu-e-disapp}. For each combination
    of data sets, we give the parameter values and the $\chi^2$ value per
    degree of freedom at the best fit point. In all fits, we treat
    $\theta_{14}$ and $\Delta m^2_{41}$ as free parameters. For the ``all
    reactor'' sample, we also leave $\theta_{13}$ free. In the
    ``$\protect\parenbar\nu_e$ disap.'' analyses, all parameters listed in
    \cref{eq:params-e-dis} are allowed to float.  For the analyses with free
    reactor fluxes, there are two additional free parameters corresponding to
    the normalization of the \iso{U}{235} and \iso{Pu}{239} fluxes. The last
    two columns of the table give the $\Delta\chi^2$ between the no-oscillation
    hypothesis and the best fit, as well as the significance at which the
    no-oscillation hypothesis is disfavoured. It is obtained by assuming that
    $\Delta\chi^2$ follows a $\chi^2$ distribution with two degrees of freedom
    ($\Delta m^2_{41}$ and $|U_{e4}|$).}
  \label{tab:reactor}
\end{table}

The new spectral data from DANSS are shown in the left panel of
\cref{fig:react-spect}.  The DANSS experiment uses a movable
detector. The plot shows the ratio of the spectra observed in two
detector locations corresponding to baselines of 10.7 and 12.7~m. The
data show a spectral distortion, leading to a preference in favour of
sterile neutrino oscillations, as illustrated by the red dashed curve in
\cref{fig:react-regions}. The remarkable observation is that the
preferred region from DANSS overlaps with the one from NEOS, which
also observes a spectral distortion consistent with sterile neutrino
oscillations, see right panel of \cref{fig:react-spect}. Results of
the combined analysis of DANSS and NEOS are given in
\cref{tab:reactor}. We find that the no-oscillation hypothesis is disfavoured with
respect to sterile neutrino oscillations at a significance of
$3.3\sigma$. Let us stress that this result is completely independent
of reactor neutrino flux predictions. It is only based on bin-by-bin
spectral comparison between two detector locations in DANSS, and
between the spectra observed in NEOS and Daya Bay.

Combing all available reactor data, we obtain the results shown
\cref{tab:reactor} and \cref{fig:react-regions}. These results confirm the
$\simeq 3\sigma$ hint in favour of sterile neutrinos from DANSS and NEOS in the
analysis with free fluxes. If the fluxes are fixed and the predicted neutrino
rate is used (``reactor anomaly''), the significance increases to $3.5\sigma$,
with a best fit point consistent with the DANSS/NEOS spectral indications. Note
that in the analysis using fixed fluxes there is minor tension between ``old''
reactor data and the DANSS/NEOS best fit region, see \cref{fig:react-regions}.
Despite this small tension, the significance for sterile neutrinos increases
from $3.3\sigma$ for NEOS+DANSS to $3.5\sigma$ for the global data. We conclude
that recent data support the indication in favour of sterile neutrinos from the
reactor anomaly, a conclusion that is solely based on spectral distortions,
but independent of reactor flux predictions.

Let us comment on the impact of the Daya Bay measurements of the
individual neutrino fluxes from different fissible
isotopes~\cite{An:2017osx} by using the time evolution of the observed
reactor anti-neutrino spectra. These data have been used to compare
the hypothesis $H_1$ of no-oscillations but free flux normalizations
to the hypothesis $H_0$ that flux predictions \cite{Mueller:2011nm,
  Huber:2011wv} (including their error estimates) are correct and a
sterile neutrino exists. Considering the test statistic
\begin{equation}
 T = \chi^2_{\rm min}(H_0)-\chi^2_{\rm min}(H_1) \,,
\end{equation}
Daya Bay data lead to $T_{\rm obs} = 6.3$, which prefers $H_1$
(flux-free) over $H_0$ (oscillations) at $2.7\sigma$
\cite{An:2017osx,Dentler:2017tkw} (see, however,
\cite{Huber:nuplatformweek}).  As shown
previously~\cite{Giunti:2017yid, Dentler:2017tkw}, this preference
decreases, once the global reactor data is combined with DayaBay
data. Using the numbers given in \cref{tab:reactor}, we find that with
present combined reactor data, $T_{\rm obs} = -1.3$, which actually
shows a slight preference for oscillations over the no-oscillation but
flux-free hypothesis. Again the main driver for this are spectral
distortions, which can be fit better by oscillations than by
re-scaling fluxes.

\subsection{Global $\protect\parenbar{\nu}_e$ Disappearance Analysis}
\label{sec:global_nu-e}

\begin{figure}
  \centering
   \includegraphics[width=.55\textwidth]{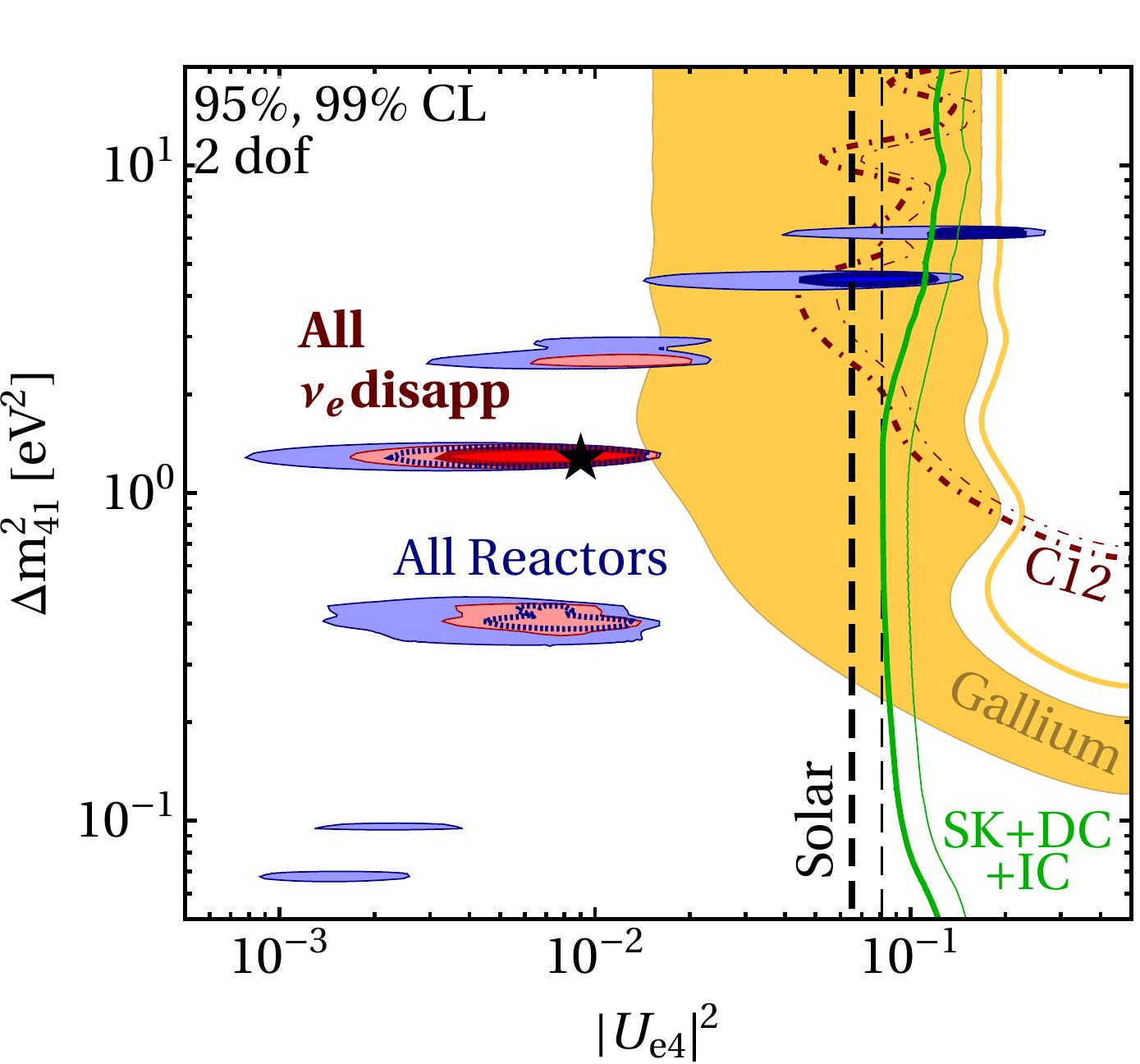}
  \caption{Constraints on $\nu_e$/$\bar\nu_e$ disappearance in the
    $3+1$ scenario.  We show the preferred parameter regions at 95\%
    and 99\%~CL, projected onto the plane spanned by the mixing matrix
    element $|U_{e4}|^2$ and the mass squared difference $\Delta
    m_{41}^2$. The parameter space inside the shaded areas and to the
    left of the exclusion curves is allowed.  For the reactor analysis
    we adopt the conservative assumption of free flux normalizations.
    The red region includes all data listed in
    \cref{tab:nu-e-disapp}. The green curves show the limit on
    $|U_{e4}|^2$ obtained from atmospheric neutrino data from SuperK,
    IceCube and DeepCore, discussed in \cref{sec:nu-mu}.}
  \label{fig:nu-e}
\end{figure}

We proceed now to combining reactor data with all other data on
$\parenbar\nu_e$ disappearance listed in \cref{tab:nu-e-disapp}. In
fitting these data we scan the following set of parameters (see
\cref{eq:U-parameterization} for our mixing matrix convention):
\begin{equation}\label{eq:params-e-dis}
  \Delta m^2_{31} , \,
  \Delta m^2_{41} , \,
  \theta_{12} ,\,
  \theta_{14} ,\,
  \theta_{24} ,\,
  \theta_{34} .
\end{equation}
We fix $\theta_{13}$ here since it is determined very
accurately, and we have checked that its best fit value does not depend on the
possible existence of sterile neutrinos~\cite{Kopp:2013vaa}.
The dependence on $\theta_{24}$ and $\theta_{34}$ appears due to solar
neutrino data, which in addition to the $\nu_e$ survival probability
includes also NC data sensitive to $\nu_e\to\nu_s$
transitions.\footnote{Formally solar neutrino data depend also on 
  complex phases~\cite{Kopp:2013vaa}. In our numerical scan we do take this effect
  into account. However, we have checked that the dependence is
  marginal and therefore we do not include phases in the counting of full
  degrees of freedom.} The
results are shown in the last two rows of \cref{tab:reactor} and in
\cref{fig:nu-e}. We observe that the best fit point remains stable at
$\Delta m^2_{41} \approx 1.3$~eV$^2$, in agreement with the
reactor-only analysis.

From \cref{fig:nu-e} we observe a slight tension
between the global best fit point and the region favoured by the
gallium anomaly. We have used the parameter goodness-of-fit (PG)
test~\cite{Maltoni:2003cu} to quantify the compatibility of the gallium
anomaly with reactor data. We obtain for the PG test-statistic (see
\cref{sec:PG} for a review) $\chi^2_\text{PG} = 4.7$, irrespective
of whether reactor fluxes are fixed or free. For 2~dof, this translates into a
$p$-value of about 9\% for the compatibility of reactors and
gallium. From \cref{fig:nu-e} we see, however, that the combined best fit point
of reactor and gallium data lies in the island around
$\Delta m^2_{41} \approx 4.5$~eV$^2$, which is disfavoured by solar
neutrinos as well as neutrino scattering on \iso{C}{12}.
For the global best fit point around
$\Delta m^2_{41} \approx 1.3$~eV$^2$, the PG test comparing reactor and gallium
data gives $\chi^2_\text{PG}
= 6.9\ (7.2)$ for fixed fluxes (free fluxes). This corresponds to a
$p$-value of 3.1\% (2.8\%), indicating some minor tension between
these data sets.
Despite this tension, \cref{tab:reactor} shows that
the significance of rejecting no-oscillations of the combined fit
increases by about two units in $\Delta\chi^2$ compared to the
reactor-only analysis, both for the flux-free and flux-fixed analyses.

In \cref{fig:nu-e} we show also the bound on $|U_{e4}|^2$ obtained
from the atmospheric neutrino experiments SuperKamiokande (SK),
IceCube (IC), and DeepCore (DC), see \cref{sec:nu-mu} for more
details. We observe that this bound is comparable to the one from
solar neutrino data. The effect of sterile neutrinos on low-energy
atmospheric data as relevant for SK and DC has been discussed in the
appendix of ref.~\cite{Maltoni:2007zf}. It amounts mostly to a
normalization effect of the electron and muon neutrino survival
probability according to $P_{\alpha\alpha} \propto (1 - 2 |U_{\alpha
  4}|^2)$ with $\alpha = e,\mu$. In our SK/DC analyses we assume a 20\%
correlated normalization error on $e$ and $\mu$-like events, and a 5\%
error on the ratio of them. Therefore, we can expect a $1\sigma$ bound
of order 0.1 on $|U_{\alpha 4}|^2$ from those data alone.  If either
$|U_{e4}|^2$ or $|U_{\mu 4}|^2$ is independently constrained from any
other data, the bound on the other one from SK/DC becomes
significantly stronger, due to the correlated uncertainty. Since the
high-energy data relevant for IC provide such an independent
constraint on $|U_{\mu 4}|^2$ due to the resonant matter effect (see
\cref{sec:nu-mu}), the combined bound improves and we get $|U_{e4}|^2
\lesssim 0.1$ at 99\%~CL (2~dof). Note that we do not include
atmospheric data in the global $\parenbar\nu_e$ disappearance analysis
presented in this section, since in this work we classify atmospheric
neutrino experiments as $\parenbar\nu_\mu$ disappearance to be
discussed below.

We conclude that global $\parenbar\nu_e$ disappearance data show a
robust hint in favour of sterile neutrinos at the $3\sigma$ level,
independent of reactor flux predictions. If reactor flux predictions
(including their uncertainties) are assumed to be correct, the
significance reaches $3.8\sigma$.

\section{$\protect\parenbar{\nu}_\mu \to \protect\parenbar{\nu}_e$ Oscillations
         at Short Baseline}
\label{sec:nu-mu-to-nu-e}

The appearance channel $\bar\nu_\mu \to \bar\nu_e$ was the first
oscillation channel to reveal possible hints for sterile neutrinos,
namely in the LSND experiment \cite{Aguilar:2001ty}.  This hint, which
to date remains the oscillation anomaly with the largest statistical
significance, was later reinforced at lower significance by
MiniBooNE~\cite{Aguilar-Arevalo:2013pmq}.  Other experiments, in
particular KARMEN~\cite{Armbruster:2002mp},
NOMAD~\cite{Astier:2003gs}, E776~\cite{Borodovsky:1992pn},
ICARUS~\cite{Antonello:2012pq, Farnese:2015kfa}, and
OPERA~\cite{Agafonova:2013xsk}, have not been able to confirm the
findings by LSND and MiniBooNE, albeit not ruling them out either.  We
summarize the data sets included in our analysis of $\nu_e$ and
$\bar\nu_e$ appearance data in \cref{tab:nu-e-app}.

\begin{table}
  \centering
  \begin{ruledtabular}
  \begin{tabular}{llcp{9cm}r}
    & Experiment   & References & Comments & Data points\\\hline
    & LSND         & \cite{Aguilar:2001ty}
                   & $\bar\nu_\mu$ from stopped pion source (DaR) & 11 \\
    & LSND         & \cite{Aguilar:2001ty}
                   & combined DaR and DiF data
                     $(\protect\parenbar\nu_\mu \to \protect\parenbar\nu_e)$ & N/A \\
    & MiniBooNE    & \cite{Aguilar-Arevalo:2013pmq, MBdataReleaseAPP}
                   & $\nu_\mu$ and $\bar\nu_\mu$ from high-energy Fermilab beam  & 22\\
    & KARMEN       & \cite{Armbruster:2002mp}
                   & $\bar\nu_\mu$ from stopped pion source & 9\\
    & NOMAD        & \cite{Astier:2003gs}
                   & $\nu_\mu$ from high-energy CERN beam & 1\\
    & E776         & \cite{Borodovsky:1992pn}
                   & $\nu_\mu$ from high-energy Brookhaven beam & 24 \\
    & ICARUS       & \cite{Antonello:2012pq, Farnese:2015kfa}
                   & $\nu_\mu$ from high-energy CERN beam & 1 \\
    & OPERA        & \cite{Agafonova:2013xsk}
                   & $\nu_\mu$ from high-energy CERN beam & 1
  \end{tabular}
  \end{ruledtabular}
  \caption{Experimental data sets included in our
    $\protect\parenbar\nu_\mu \to \protect\parenbar\nu_e$ analysis. For LSND,
    we have carried out analyses using only decay-at-rest (DaR) data, or the
    combination with decay-in-flight (DiF) data. In the latter case we use a
    $\chi^2$ table provided by the collaboration, which cannot be associated
    with a number of data points. The total number of data points in the
    appearance channel (when using LSND DaR data only) is 69.
  }
  \label{tab:nu-e-app}
\end{table}

Compared to our previous publication, ref.~\cite{Kopp:2013vaa}, in which
more technical details on our fits are given, we have added the following
data sets:
\begin{enumerate}
  \item New results from the ICARUS
    \cite{Antonello:2012pq, Farnese:2015kfa} and OPERA \cite{Agafonova:2013xsk}
    experiments in the high energy ($\sim 20$\,GeV) CNGS beam. Both experiments
    have searched for anomalous $\nu_\mu \to \nu_e$ appearance, but have not
    found any evidence. They are thus able to impose constraints over a wide range
    of $\Delta m_{41}^2$ values.

  \item Decay-in-flight data from LSND. The neutrino oscillation
    analysis of LSND is based on a search for anomalous $\bar\nu_e$ appearance
    in the neutrino flux from a stopped pion source. Since the LSND detector
    was placed downstream from the pion production target, it received not only
    $\nu_\mu$, $\bar\nu_\mu$, and $\nu_e$ from $\pi^+$ decays at rest (DaR),
    but also neutrinos and anti-neutrinos from pions decaying in flight (DiF).
    A discussion of the impact of DiF data in the context of the global sterile
    neutrino fit can be found in ref.~\cite{Maltoni:2002xd}. The LSND
    collaboration has kindly provided tabulated $\chi^2$ values from their
    combined DaR+DiF fit. The LSND fit is based on the two-flavour
    approximation, so to include the tabulated $\chi^2$ values in our 4-flavour
    analysis, we compute at each parameter point the effective two-flavour
    mixing angle
    \begin{align}
      \sin^2 2\theta_{\mu e} \equiv 4 |U_{e4}|^2 |U_{\mu 4}|^2 \,.
      \label{s22thme}
    \end{align}
    from the full four-flavour mixing matrix $U$.
    In the following, we will show results using both our
    previous fitting code that includes only DaR data as well as results
    based on the tabulated two-flavour $\chi^2$ values from
    LSND for DaR+DiF data.
\end{enumerate}

\begin{figure}
  \centering
  \begin{tabular}{cc}
    \includegraphics[width=.48\textwidth]{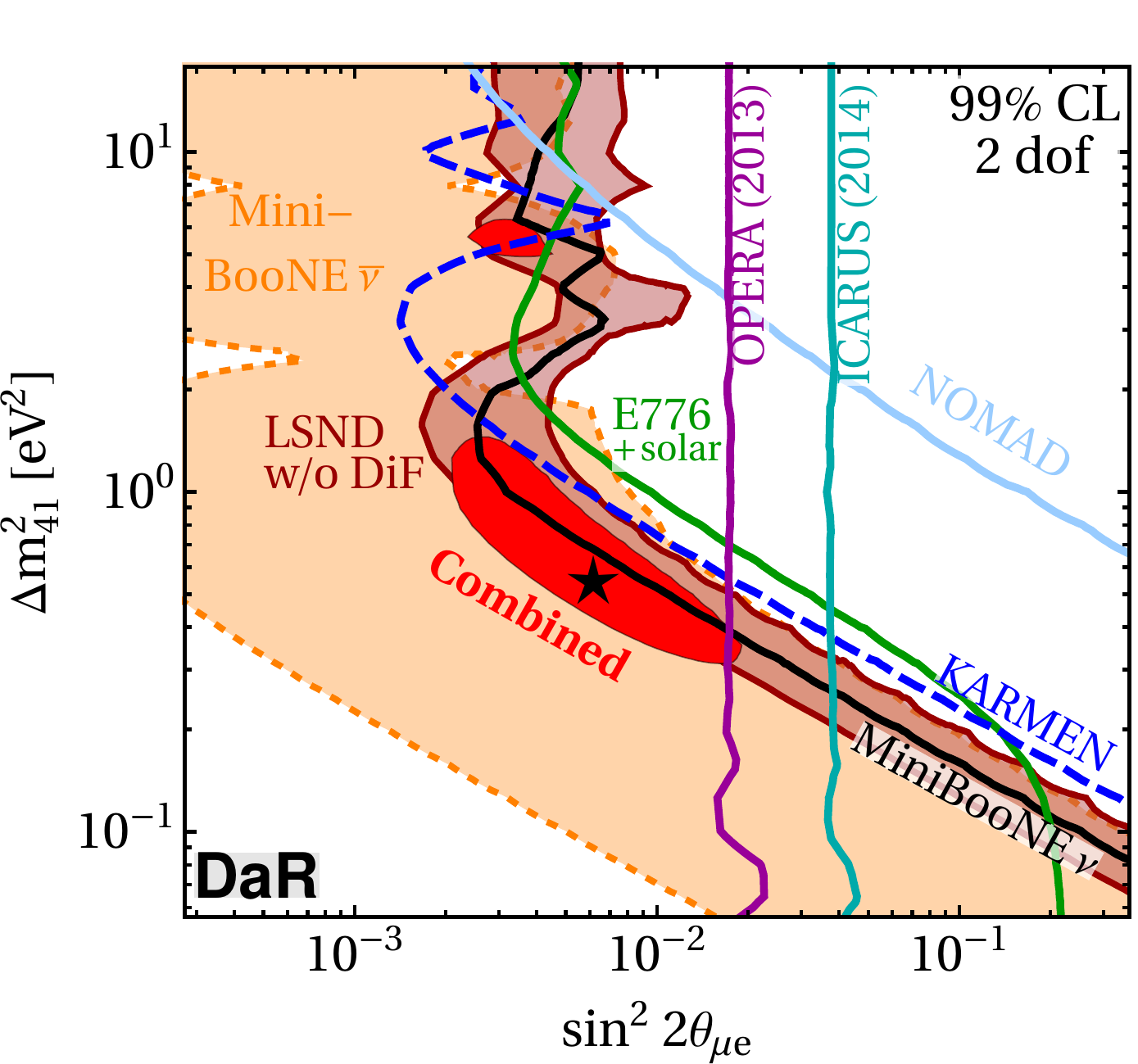}
    \includegraphics[width=.48\textwidth]{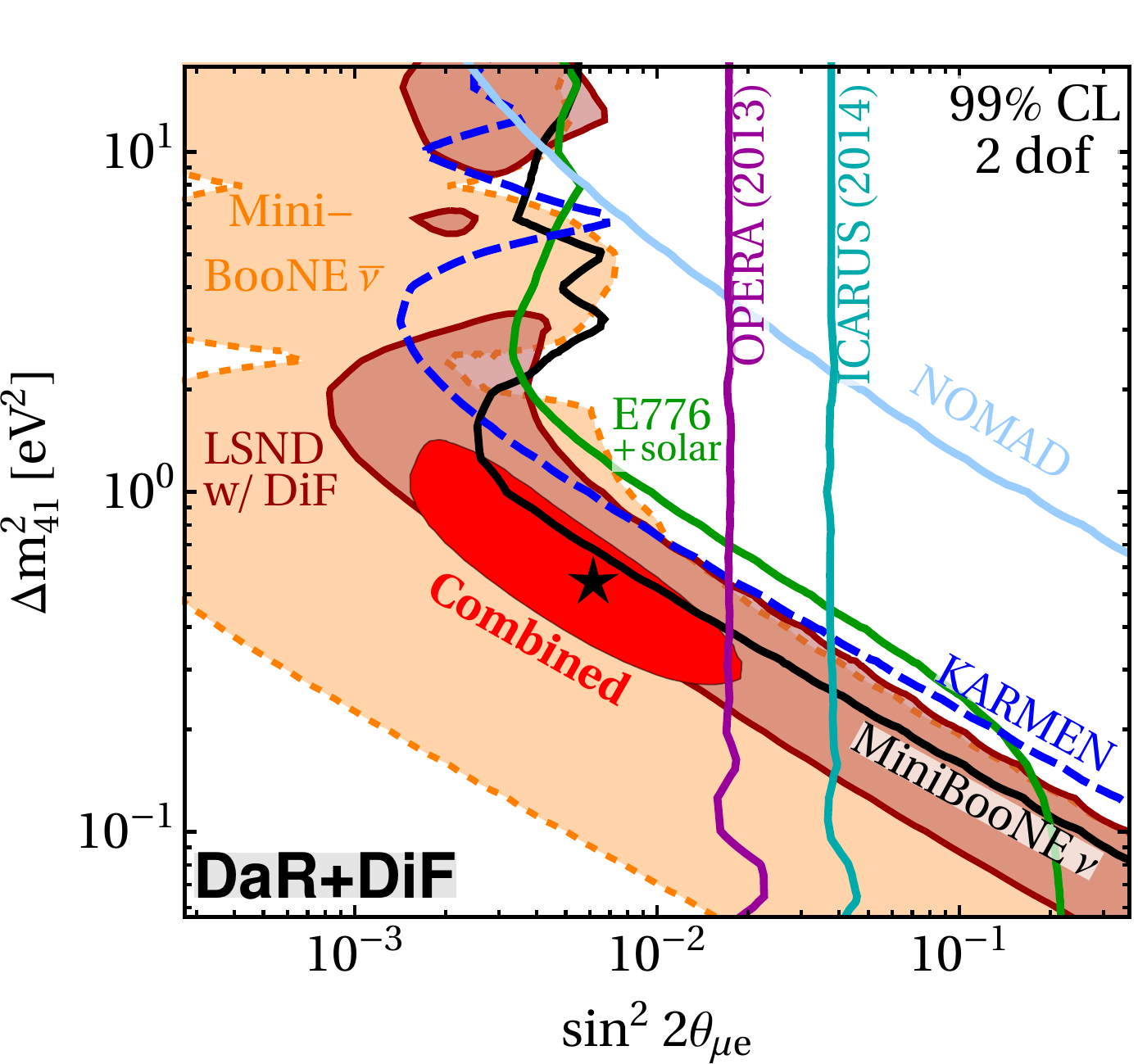}
  \end{tabular}
  \caption{Constraints on short-baseline $\nu_\mu \to \nu_e$ and $\bar\nu_\mu
    \to \bar\nu_e$ oscillations in the presence of sterile neutrinos in $3+1$
    scenarios.  We show the allowed parameter regions, projected onto the plane
    spanned by the effective mixing angle $\sin^2 2\theta_{\mu e} \equiv 4
    |U_{e4}|^2 |U_{\mu 4}|^2$ and the mass squared difference $\Delta
    m_{41}^2$. In the left panel only decay-at-rest (DaR) data from LSND is
    included, while in the right panel also decay-in-flight data (DiF) is
    used.}
  \label{fig:nu-mu-to-nu-e}
\end{figure}

Our results are plotted in \cref{fig:nu-mu-to-nu-e}, which shows the favoured
parameter regions projected onto the $\sin^2 2\theta_{\mu e}$--$\Delta
m_{41}^2$ plane.  We see that all $\parenbar\nu_\mu \to \parenbar\nu_e$ data
sets are consistent among each other: a large chunk of the parameter region
favoured by LSND and MiniBooNE is not probed by any of the other searches.  The
strongest constraints come from OPERA at $\Delta m_{41}^2 \lesssim 0.5$~eV$^2$,
and from KARMEN at larger $\Delta m_{41}^2$. Note that data from E776 is
combined with solar neutrino data because a fit to E776 data alone would not be
meaningful as it would leave possible oscillations of the $\nu_e$ and
$\bar\nu_e$ backgrounds into sterile states unconstrained. Fitting E776 data
jointly with solar neutrino data provides a reasonable constraint on
$|U_{e4}|$, cf.\ \cref{fig:nu-e}.

The conclusions drawn from \cref{fig:nu-mu-to-nu-e} agree qualitatively with
the ones from our earlier paper ref.~\cite{Kopp:2013vaa}. Some constraints, in
particular those from OPERA and ICARUS, have become significantly stronger and
now disfavour values of $\sin^2 2\theta_{\mu e} \gtrsim 0.02$ that were still
allowed previously. Note that our OPERA and ICARUS limits deviate slightly from
those published by the respective collaborations~\cite{Antonello:2012pq,
Farnese:2015kfa, Agafonova:2013xsk} because we include oscillations of the
backgrounds. Moreover, for consistency with the other exclusion curves in
\cref{fig:nu-mu-to-nu-e}, we interpret the $\chi^2$ values from our OPERA and
ICARUS fits assuming two degrees of freedom.  We have checked that our code
reproduces the official limits from refs.~\cite{Antonello:2012pq,
Farnese:2015kfa, Agafonova:2013xsk} very well when the same assumptions as in
the official publications are used.

Let us mention that the global $\parenbar\nu_\mu \to \parenbar\nu_e$
analysis has a relatively poor goodness of fit. For the combined best
fit point using the LSND DaR analysis we find
$\chi^2_\text{min}/\text{dof} = 89.9 / (69-2)$, which corresponds to a
$p$-value of 3.3\%. This is mostly driven by the MiniBooNE low-energy
excess, which cannot be fitted well in the $3+1$ scenario, and by
the contribution from E776 whose spectrum gives a relatively poor
fit. This feature has been present also in our previous analysis
\cite{Kopp:2013vaa}, where a more detailed discussion can be found.

In all cases LSND dominates the appearance fit. LSND alone disfavours
the no-oscillation hypothesis with $\Delta\chi^2 = 44\ (29)$ when
using DaR (DaR+DiF) data. For the combined appearance analysis these
numbers increase slightly, due to the hint for appearance in MiniBooNE
data. We find that the no-oscillation hypothesis for all appearance
data is disfavoured compared to the best fit by $\Delta\chi^2 =
46\ (35)$ when using LSND DaR (DaR+DiF) data.

Comparing the allowed regions with and without the inclusion of decay-in-flight
data in LSND, we see that the impact on the global fit is relatively
minor.  This is because although the LSND region with DiF data extends
to slightly smaller values of $\sin^2 2\theta_{\mu e}$, MiniBooNE
appearance data prefers smaller $\Delta m_{41}^2$ and mixing angles
(especially for the neutrino mode data), somewhat limiting the impact
of LSND DiF data when LSND and MiniBooNE data are combined.  We observe
only a slight broadening of the parameter regions preferred by LSND
and by the combination of all $\nu_\mu \to \nu_e$ and $\bar\nu_\mu \to
\bar\nu_e$ appearance data.  We will see in \cref{sec:gof} that this
slightly reduces the tension between appearance and disappearance
data, but does not remove it.

\section{$\protect\parenbar{\nu}_\mu$ Disappearance Data}
\label{sec:nu-mu}

Searches for muon neutrino disappearance due to oscillations involving a fourth
neutrino mass state have recently received a significant boost thanks to novel
results on sterile neutrinos from atmospheric neutrino data (both in the TeV
energy window from IceCube~\cite{TheIceCube:2016oqi} and at lower energy from
DeepCore~\cite{Aartsen:2014yll}) as well as from a combined analysis of MINOS
and MINOS+ charged current (CC) and neutral current (NC) data
\cite{Adamson:2017uda}.  Also NO$\nu$A has presented a first search for sterile
neutrinos based on NC data~\cite{Adamson:2017zcg}. Searches for a deficit of NC
events are of particular interest because they are sensitive to mixing of
sterile neutrinos with \emph{any} active neutrino flavor. As such, any deficit
found would be a unique signature of sterile neutrinos.  The new analyses by
IceCube, DeepCore, MINOS/MINOS+, and NO$\nu$A complement, and significantly
extend, the exclusion regions from the short-baseline experiments
CDHS~\cite{Dydak:1983zq} and MiniBooNE~\cite{AguilarArevalo:2009yj,
Cheng:2012yy}, from Super-Kamiokande data on atmospheric
neutrinos~\cite{Wendell:2010md, wendell:2014dka}, and from
MINOS~\cite{MINOS:2016viw}.

The high-energy IceCube analysis from ref.~\cite{TheIceCube:2016oqi} exploits the fact
that active-to-sterile neutrino oscillations in matter are resonantly enhanced
by the MSW effect~\cite{Wolfenstein:1977ue, Mikheev:1986gs} at an energy of
\begin{align}
  E_\text{res} = 5.3~\text{TeV} \times
                 \bigg( \frac{5~ \text{g}/\text{cm}^3}{\rho_\oplus} \bigg)
                 \bigg( \frac{\Delta m_{41}^2}{1~\text{eV}^2} \bigg) \,.
  \label{eq:icecube-resonance}
\end{align}
Here $\rho_\oplus$ is the mass density of the material through which
neutrinos are propagating. It is on average $\sim
3~\text{g}/\text{cm}^3$ in the Earth's crust and outer mantle, $\sim
5~\text{g}/\text{cm}^3$ in the inner mantle, and between 10 and
$13~\text{g}/\text{cm}^3$ in the
core~\cite{Dziewonski:1981xy}. \Cref{eq:icecube-resonance} implies
that, for sterile neutrinos at the eV-scale, neutrino telescopes like
IceCube can in principle observe maximal oscillations at TeV energies
--- a sweet spot well above the detection threshold, but still low
enough for the atmospheric neutrino flux to be
appreciable~\cite{Nunokawa:2003ep, Choubey:2007ji}. For larger or
smaller $\Delta m_{41}^2$, the sensitivity is expected to dwindle as
the resonance moves to energies with a lower neutrino flux, or moves
below the energy threshold of the detector.  A limiting factor to this
analysis is the fact that, for $\Delta m_{41}^2 > 0$ as considered
here, the resonance is in the anti-neutrino sector. Since neutrino
telescopes cannot distinguish neutrinos from anti-neutrinos on an
event-by-event basis, and since anti-neutrino cross sections are
smaller by about a factor of three than neutrino cross sections, the
magnitude of the observable effect is reduced.\footnote{For $\Delta
  m_{41}^2 < 0$ the resonance would occur for neutrinos and the signal
  would therefore be stronger. However, such scenarios are in strong
  tension with cosmology.}  Moreover, for small mixing angles, the
resonance width,
\begin{align}
  \Delta E_\text{res} \sim \frac{\Delta m_{41}^2 \sin^2 2\theta_{24}}{2 V_\text{MSW}} \,,
  \label{eq:ic-res-width}
\end{align}
is small, so that only a very small fraction of the energy spectrum is
affected. The narrow width, combined with the limited experimental energy resolution,
further reduces the sensitivity of IceCube.
In \cref{eq:ic-res-width}, $V_\text{MSW} \simeq 1.9 \times
10^{-14}~\text{eV} \times [\rho_\oplus / (\text{g}/\text{cm}^3)]$ is the
neutral current-induced MSW potential for muon and tau neutrinos.
Finally, systematic uncertainties play a crucial role in the analysis
from ref.~\cite{TheIceCube:2016oqi}.
Technical details on our implementation of the IceCube analysis are given in
\cref{sec:icecube-fit}.

In addition to the TeV neutrino events discussed above, the IceCube
collaboration has also observed atmospheric neutrinos in the
tens-of-GeV range through its sub-detector DeepCore. The information
on sterile neutrinos which can be extracted from this low-energy
sample is very similar to that provided by Super-Kamiokande
atmospheric data, which has been discussed in detail in
refs.~\cite{Maltoni:2007zf, Kopp:2013vaa}. As explained there,
low-energy atmospheric neutrino data can put a strong bound on $|U_{\mu 4}|^2$
through the suppression of the $P_{\mu\mu}$ oscillation probability which a mixing
of $\nu_\mu$ with a heavy state would imply. Moreover, such data also
constrains $|U_{\tau 4}|^2$ because the zenith-angle
dependence of $P_{\mu\mu}$ is modified if oscillations driven by
$\Delta m_{31}^2$ deviate from vacuum-like $\nu_\mu \to \nu_\tau$
oscillations. The formalism for neutrino oscillations discussed in appendix~D
of ref.~\cite{Kopp:2013vaa} for Super-Kamiokande phase~1--3 data is also
applied here to phase~4 results as well as to DeepCore data.

The MINOS detector is particularly interesting for sterile neutrino searches as
it has observed neutrino oscillations over a fairly wide range of energies:
during the original MINOS run, the NuMI beam was tuned to a peak energy of
$\sim 2$~GeV, while in the MINOS+ phase, the peak energy was at about 6~GeV,
with the spectrum extending to tens of GeV.  Moreover, the MINOS collaboration
has analysed not only CC $\nu_\mu$ disappearance sensitive mainly to $U_{\mu
4}$, but has also searched for disappearance in NC events.  Since MINOS/MINOS+
has near and far detectors, the experiment is sensitive over a wide range of
$\Delta m_{41}^2$ values. For $\Delta m_{41}^2 \sim
\text{$10^{-3}$--$10^{-1}$~eV}^2$, an oscillation pattern can be observed in
the far detector, while no oscillations are expected in the near detector.  At
larger mass squared difference, oscillations in the far detector enter the
averaging regime. At $\Delta m_{41}^2 \sim \text{1--100~eV}^2$, oscillation
patterns begin to emerge in the near detector.  In our analysis of MINOS/MINOS+
data, we follow very closely the recommendations accompanying the MINOS/MINOS+
data release \cite{Adamson:2017uda}.

We have also implemented the NO$\nu$A neutral current analysis from
ref.~\cite{Adamson:2017zcg}.  Due to the low number of events and the difficult
reconstruction of the neutrino energy in NC events, only total rates are used
in the analysis.  The dominant background in this analysis are misidentified
charged current events.  Following ref.~\cite{Adamson:2017zcg}, we implement a
12.2\% (15.3\%) systematic uncertainty on the signal (background) rates.
Compared to the MINOS/MINOS+ NC search, the narrow-band beam employed in
NO$\nu$A means that the experiment is sensitive to a much smaller range of
$\Delta m^2_{41}$ values, namely between $0.05~\eV^2$ and $0.5~\eV^2$.  Even in
this mass range, the NO$\nu$A search for sterile neutrinos is not competitive
with other searches yet as it is suffers from large systematic uncertainties
related to detector modelling and energy reconstruction, but it is expected to
improve considerably in the future. 

\begin{table}
  \centering
  \begin{ruledtabular}
  \begin{tabular}{llcp{7cm}r}
    & Experiment   & References & Comments & Data points\\ \hline
    & IceCube (IC) & \cite{TheIceCube:2016oqi,Jones:2015, Arguelles:2015}
                   & MSW resonance in high-$E$ atmospheric $\bar\nu_\mu$  & 189\\
    & CDHS         & \cite{Dydak:1983zq}
                   & accelerator $\nu_\mu$ & 15\\
    & MiniBooNE    & \cite{AguilarArevalo:2009yj, Cheng:2012yy, MBdataReleaseDIS}
                   & accelerator $\nu_\mu$ and $\bar\nu_\mu$ & 15 + 42\\
    & Super-Kamiokande (SK) & \cite{Wendell:2010md, wendell:2014dka}
                       & low-$E$ atmospheric neutrinos & 70\\
    & DeepCore (DC)    & \cite{Aartsen:2014yll, deepcore:2016}
                       & low-$E$ atmospheric neutrinos & 64\\    
    & NO$\nu$A     & \cite{Adamson:2017zcg}
                   & NC data & 1 \\
    & MINOS/MINOS+ & \cite{Adamson:2017uda}
                   & accelerator $\nu_\mu$, CC \& NC event spectra & 108
  \end{tabular}
  \end{ruledtabular}
  \caption{Experimental data sets included in our
    $\protect\parenbar\nu_\mu \to \protect\parenbar\nu_\mu$
    disappearance analysis. The total number of data points in this channel is 504.
  }
  \label{tab:nu-mu-disapp}
\end{table}

\begin{figure}
  \centering
  \includegraphics[width=.55\textwidth]{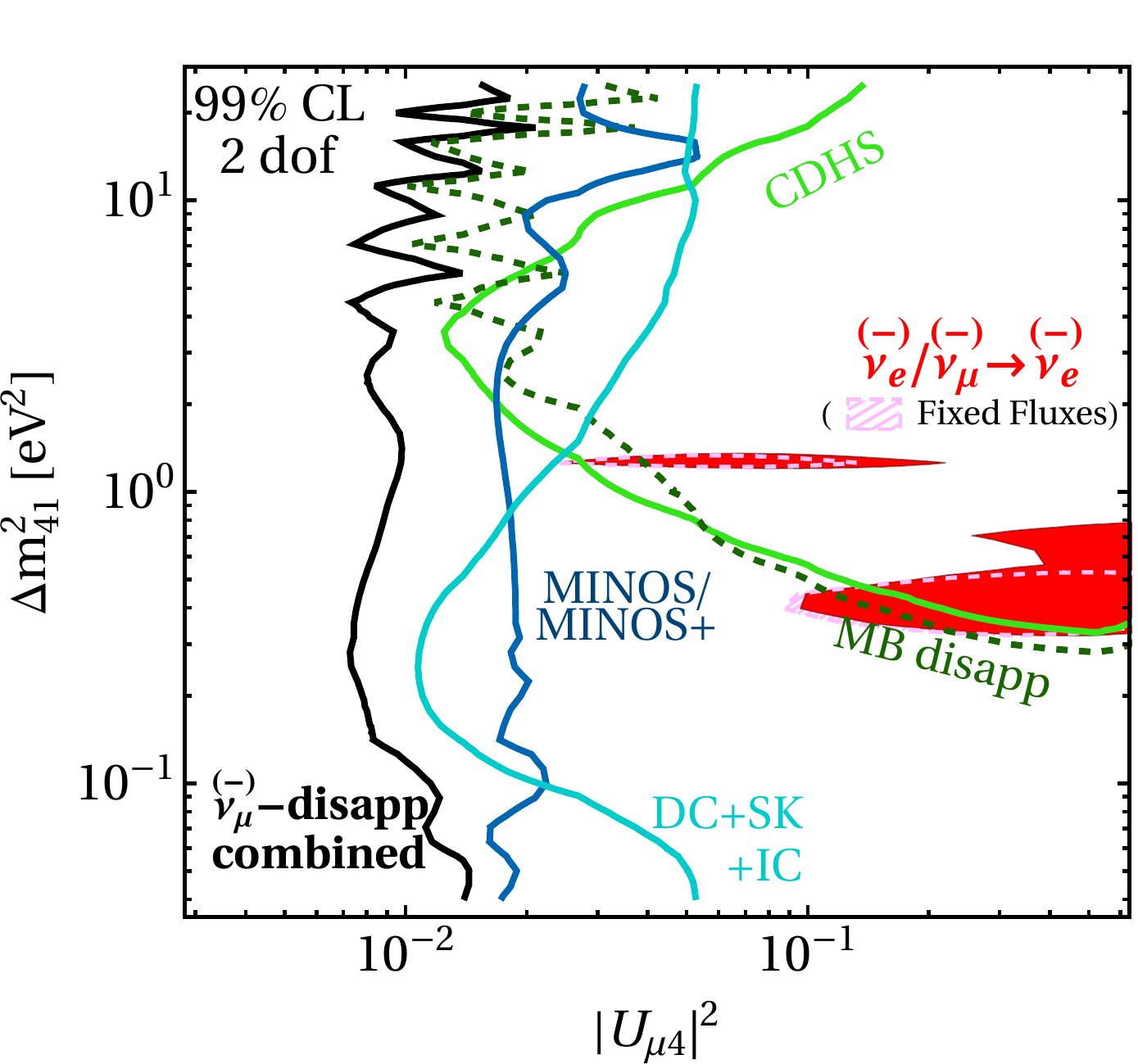}
  \caption{Constraints on the $3+1$ scenario from
    $\nu_\mu$/$\bar\nu_\mu$ disappearance.  We show the allowed
    parameter regions, projected onto the plane spanned by the mixing
    matrix element $|U_{\mu 4}|^2$ and the mass squared difference
    $\Delta m_{41}^2$. Note that the exclusion limit from NO$\nu$A is
    still too weak to appear in the plot. It is, however, included in
    the curve labelled ``combined'', which includes all data listed in
    \cref{tab:nu-mu-disapp}. The curve labelled DC+SK+IC combines all
    our atmospheric neutrino data; for this bound we have fixed the
    parameters $\theta_{12},\theta_{13},\theta_{14}$ but minimize with
    respect to all other mixing parameters, including complex phases.
    For comparison, we also show the parameter region favoured by
    $\nu_e$ disappearance and $\nu_\mu \to \nu_e$ appearance data
    (using LSND DaR+DiF), projected onto the $|U_{\mu 4}|^2$--$\Delta
    m_{41}^2$ plane; we show the allowed regions for the analyses with
    fixed and free reactor neutrino fluxes.}
  \label{fig:nu-mu}
\end{figure}

We summarize the $\nu_\mu$/$\bar\nu_\mu$ disappearance data sets
included in our analysis in \cref{tab:nu-mu-disapp}.  Details on the
CDHS and MiniBooNE analyses are given in ref.~\cite{Kopp:2013vaa} and
in the references therein.  Our results are shown in \cref{fig:nu-mu}
as a function of the mixing matrix element $|U_{\mu 4}|^2$ and the
mass squared difference $\Delta m_{41}^2$. The plot reveals strong limits
of order $|U_{\mu 4}|^2 \lesssim 10^{-2}$ across a wide range of $\Delta
m_{41}^2$ values from $\sim 2 \times 10^{-1}\,\text{eV}^2$ to $\sim
10\,\text{eV}^2$.  MINOS/MINOS+ gives an important contribution in
most of the parameter space.  The strong constraint from atmospheric
neutrino data at $\Delta m_{41}^2 \lesssim 1\,\text{eV}^2$ is
dominated by IceCube.  At large masses, MiniBooNE and to some extent CDHS
are competitive with the MINOS/MINOS+ bound.  Comparing to the parameter
region preferred by appearance and $\nu_e$/$\bar\nu_e$ disappearance
data (which includes the oscillation anomalies), we see dramatic
tension.  Given the constraints on $U_{e4}$ from reactor experiments,
the values of $\sin^2 2\theta_{\mu e} \equiv 4 |U_{e4}|^2 |U_{\mu
  4}|^2$ required by LSND and MiniBooNE can only be reached if
$|U_{\mu 4}|$ is large.  This, however, is clearly disfavoured by
multiple $\nu_\mu$/$\bar\nu_\mu$ disappearance null results. This is
the origin of the severe tension in the global fit we are going to
report below. As we are going to discuss, this tension has become very
robust and does not rely on any single $\parenbar\nu_\mu$
disappearance data set.

\section{Constraints on $|U_{\tau 4}|$}
\label{sec:U-tau-4}

Mixing between tau neutrinos and possible sterile states is
particularly difficult to constrain since no $\nu_\tau$ sources are
available. Nevertheless, constraints can be obtained in the following
two ways: \emph{(i)} studying matter effects. All active neutrino
flavors experience an MSW potential caused by coherent forward
scattering through $Z$ boson exchange, while sterile neutrinos do
not. This influences $\nu_e$ disappearance observed in solar neutrino
experiments, as well as $\nu_\mu$ disappearance observed in beam
experiments and in atmospheric neutrinos. The latter yield
particularly strong limits as they possess the longest baselines in
matter.  \emph{(ii)} exploiting neutral current events, which are
sensitive to \emph{any} disappearance of active neutrinos. This
approach allows us to derive constraints from the sterile neutrino
searches in MINOS/MINOS+~\cite{Adamson:2017uda} and
NO$\nu$A~\cite{Adamson:2017zcg}, and from SNO solar neutrino
data~\cite{Aharmim:2007nv, Aharmim:2005gt, Aharmim:2008kc}. The
corresponding analysis codes used in our fit are the same as discussed
in \cref{sec:nu-e,sec:nu-mu}. Compared to ref.~\cite{Kopp:2013vaa}, we
have in particular added IceCube, DeepCore, MINOS/MINOS+, and NO$\nu$A
data to the fit.

\begin{figure}
  \centering
  \includegraphics[width=\textwidth]{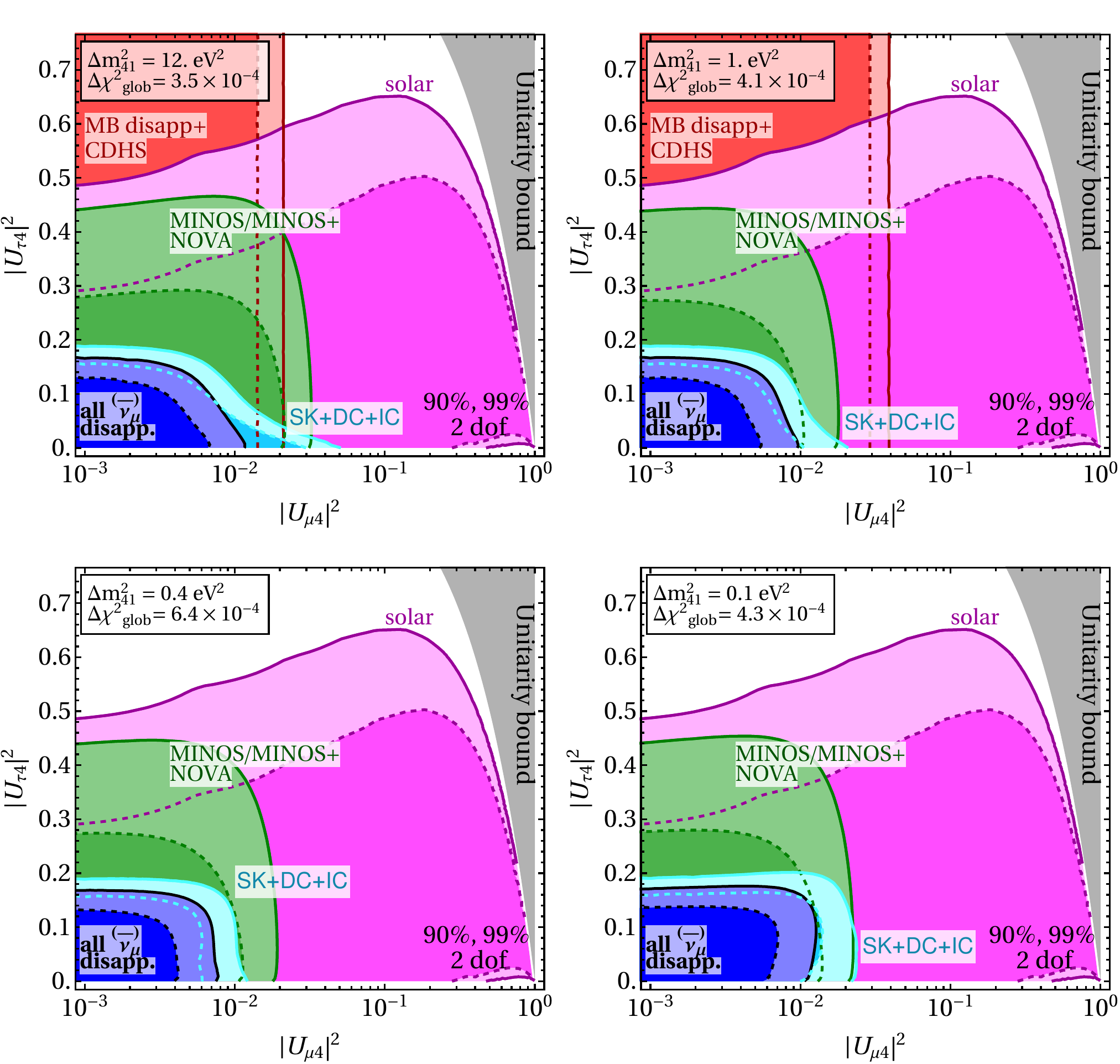}
  \caption{Constraints on the mixing of sterile neutrinos with muon and tau
    neutrinos, parameterized by the corresponding elements $|U_{\mu 4}|$
    and $|U_{\tau 4}|$ of the leptonic mixing matrix. In each panel, $\Delta m_{41}^2$
    has been fixed to a different value, while $\Delta m_{31}^2$, $\theta_{23}$,
    $\theta_{12}$ and $\theta_{14}$, as well as complex phases have been profiled out
    in those experiments where they have a significant impact.
    Exclusion contours are drawn relative to the minimum $\chi^2$ in each panel;
    the difference to the global minimum $\chi^2$ is indicated in each plot.
    Grayed out areas show the parameter region incompatible with the unitarity of
    the leptonic mixing matrix.}
  \label{fig:Umu4-Utau4}
\end{figure}

Our results are shown in the four panels of
\cref{fig:Umu4-Utau4}. Each panel corresponds to a different fixed
value of $\Delta m_{41}^2$, and the corresponding contours have been
drawn based on the $\chi^2$ differences relative to the best fit point
for this fixed $\Delta m_{41}^2$.  The difference in $\chi^2$ between
the individual best fit points and the global one are, however, very
small, as indicated in each panel. The reason is that in all cases the
best fit point is very close to zero mixing, and therefore has very
similar $\chi^2$ values.  In defining the exclusion contours we have
assumed a $\chi^2$ distribution with two degrees of freedom. We see
that depending on $\Delta m_{41}^2$, the limit on $|U_{\mu 4}|$ is
driven by MINOS/MINOS+, IceCube, or the short-baseline experiments
MiniBooNE and CDHS, in agreement with \cref{fig:nu-mu}.  The strongest
constraints on $|U_{\tau 4}|$ typically come from atmospheric
neutrinos. We find that the combined bound is independent of $\Delta
m^2_{41}$ and is given by
\begin{align}
  |U_{\tau 4}|^2 < 0.13 \,(0.17) \quad\text{at}\quad 90\% \,(99\%) \, \text{CL}. 
\end{align}

Let us mention that recently ref.~\cite{Blennow:2018hto} has found a $2\sigma$
hint from Ice Cube data in favour of sterile neutrinos with non-zero
$\nu_4$--$\nu_\tau$ mixing in the high-mass region, with $\Delta m^2_{41}
\simeq 100$~eV$^2$. With our code we cannot reproduce their results and we do
not find any hint for sterile neutrino mixing in that mass range. The origin of
these different results is currently under investigation.

\section{The Disappearance--Appearance Tension}
\label{sec:gof}

As discussed above, results on the $\nu_e\to\nu_e$,
$\nu_\mu\to\nu_e$, and $\nu_\mu\to\nu_\mu$ oscillation channels (and
the corresponding anti-neutrino modes) over-constrain eV-scale sterile
neutrino models.  The reason can be easily understood by going to the
short-baseline limit in which baselines are so short that oscillations
induced by $\Delta m^2_{31}$ and $\Delta m^2_{21}$ did
not yet develop. In this limit, \cref{eq:P-SBL-dis,eq:P-SBL-app} show that the
bounds on $|U_{e4}|$ and $|U_{\mu 4}|$ from electron and muon
disappearance data lead to a quadratic suppression of the effective
amplitude $\sin^22\theta_{e\mu}$, \cref{s22thme}, relevant for
$\nu_\mu\to\nu_e$ appearance \cite{Bilenky:1996rw, Okada:1996kw,
  Barger:1998bn}. Thus constraints from disappearance data
challenge an explanation of the anomalies in the appearance channel in
terms of sterile neutrino oscillations. While this tension has persisted
for a very long time, see for instance ref.~\cite{Maltoni:2002xd},
it has become exceedingly severe with recent data, rendering the
sterile neutrino hypothesis as an explanation for the appearance
anomalies very unlikely, see below.

\begin{table}
  \centering
  \begin{tabular}{l@{\quad}l@{\quad}r@{\quad}l}
    \hline\hline
    Data set & Reference & Data points & Relevant parameters \\
    \hline
    $\protect\parenbar\nu_e$  disappearance 
      & \Cref{tab:nu-e-disapp}  &  594  & $\Delta m^2_{31}$,
                                          $\Delta m^2_{41}$,
                                          $\theta_{12}$,
                                          $\theta_{14}$,
                                          $\theta_{24}$,
                                          $\theta_{34}$ \\
    $\protect\parenbar\nu_\mu$ disappearance
      & \Cref{tab:nu-mu-disapp} &  504  & $\Delta m^2_{31}$,
                                          $\Delta m^2_{41}$,
                                          $\theta_{23}$,
                                          $\theta_{14}$,
                                          $\theta_{24}$,
                                          $\theta_{34}$ \\
    $\protect\parenbar\nu_\mu \to \parenbar\nu_e$ appearance (w/o LSND DiF)
      & \Cref{tab:nu-e-app}     &   69  & $\Delta m^2_{41}$,
                                          $|U_{e4}U_{\mu 4}|$ \\
    \hline
    \multicolumn{2}{l}{Total number of data points:}                   & 1167 \\
    \hline\hline
  \end{tabular}
  \caption{Number of degrees of freedom and parameters relevant to the
    counting of degrees of freedom for each data set. More details on
    the individual experiments are given in the corresponding
    tables. The number of degrees of freedom for the LSND
    decay-in-flight analysis is not available.  Thus, in the sum of
    degrees of freedoms for appearance and all data sets, we used the
    LSND decay-at-rest number. See text for details and comments on
    additional nuisance parameters.}
  \label{tab:dofs}
\end{table}

\begin{table}
  \centering
  \begin{ruledtabular}
\begin{small}
  \begin{tabular}{lccccccc}
    Analysis                       & $\Delta m_{41}^2$ [eV$^2$] & $|U_{e4}|$ & $|U_{\mu 4}|$ & $\chi^2_\text{min}/\text{dof}$ & GOF
                                   & $\chi^2_\text{PG}$ & PG \\
    \hline
    appearance (DaR)       &  0.573  & 
\multicolumn{2}{c}{$4 |U_{e4}|^2 U_{\mu 4}|^2 = 6.97\times10^{-3}$}        
            & 89.8/67  & 3.3\% \\
    appearance (DiF)       & {0.559}                   & 
\multicolumn{2}{c}{$4 |U_{e4}|^2 U_{\mu 4}|^2 = { 6.31\times10^{-3} }$} & {79.1/$-$}  \\
    $\protect\parenbar\nu_\mu$ disapp & {$2\times10^{-3}$} & {$0.12$} & {$0.039$}  & {468.9/497} & 81\% \\
    \hline
    \multicolumn{5}{l}{\bf Reactor fluxes fixed at predicted value $\pm$ quoted uncertainties} \\
    $\protect\parenbar\nu_e$ disapp     & {1.3}  & {0.1}   & $-$ 
    & 552.8/588  & 85\% \\    
    Global (DiF) & {6.03} & {0.2}  & {0.1}  & {1127/$-$} &       & 25.7 & {$2.6\times10^{-6}$} \\
    Global (DaR) & {5.99} & {0.21} & {0.12} & {1141/1159} & 64\% & 28.9 & {$5.3\times10^{-7}$} \\
    \hline
    \multicolumn{5}{l}{\bf Reactor fluxes floating freely} \\
    $\protect\parenbar\nu_e$ disapp     & {1.3} & {0.095}  & $-$ 
    & 542.9/586 & 90\% \\
    Global (DiF) & {6.1} & {0.20} & {0.10} & 1121/$-$   &      & 29.6 & $3.7\times10^{-7}$ \\
    Global (DaR) & {6.0} & {0.22} & {0.11} & 1134/1157  & 68\% & 32.1 & $1.1\times10^{-7}$ \\
  \end{tabular}
\end{small}
  \end{ruledtabular}
  \caption{Parameter values at the global best fit point and at the
    best fit points obtained for subsets of the data. We also indicate
    the $\chi^2$ per degree of freedom at the best fit points, as well
    as the corresponding goodness-of-fit values. The numbers of data
    points, and the parameters relevant to the counting of degrees of freedom
    are summarized in \cref{tab:dofs}.  For
    the global fit, we also indicate the results of the parameter
    goodness-of-fit test \cite{Maltoni:2003cu} comparing appearance to
    disappearance data. The labels ``DaR'' and ``DiF'' refer to the
    LSND analysis employed, where ``DiF'' implies the joint use of DaR+DiF
    data, see \cref{sec:nu-mu-to-nu-e}.  Note that, as the number of
    degrees of freedom for the LSND DiF data is not available, we do
    not list the corresponding goodness of fit values.}
  \label{tab:global-bf}
\end{table}

The results of the combined fit are summarized in
\cref{tab:global-bf}, which shows the results for $\parenbar\nu_e$
disappearance, $\parenbar\nu_\mu$ disappearance, and $\parenbar\nu_e$
appearance data separately as well as combined. The total numbers of
data points in these analyses are summarized in \cref{tab:dofs}. The
last column of that table also indicates which parameters need to be
considered when counting degrees of freedom.  For the
$\parenbar\nu_\mu$ disappearance data we do take into account complex
phases in the fit~\cite{Kopp:2013vaa}, but since numerically their
effect is very small we do not count them as full dof. We do, however,
treat the normalization of the atmospheric neutrino flux as a free
parameter in the IceCube analysis. Concerning the appearance sample,
for most of the data summarized in \cref{tab:nu-e-app} the
short-baseline approximation holds, motivating the use of only the
effective mixing angle quoted in \cref{tab:dofs}.  Exceptions are the
long-baseline experiments ICARUS and OPERA, which depend on more
parameters, but play a role neither for the appearance best fit point
nor for the global best fit point. Therefore, we consider only two
effective parameters for the appearance sample. For the global
analysis we count seven parameters plus the IceCube global
normalization. The reactor analysis with free fluxes has two
additional free parameters.

We would now like to quantify the tension between different subsets of
the global data that is evident from \cref{fig:nu-mu}.  We first note
that combining all data sets we find a goodness-of-fit for the global
best fit point around 65\%, see \cref{tab:global-bf}.  This good
$p$-value does not reflect the tension we found because many data
points entering the global fit have only little sensitivity to sterile
neutrino oscillations, thus diluting the power of a goodness-of-fit
test based on $\chi^2/\text{dof}$.

A more reliable method for quantifying the compatibility of different
data sets is the parameter goodness-of-fit (PG)
test~\cite{Maltoni:2003cu}, which measures the penalty in $\chi^2$
that one has to pay for combining data sets, see \cref{sec:PG} for a
brief review of this test. If the global neutrino oscillation data
were consistent when interpreted in the framework of a $3+1$ model,
\emph{any} slicing into two statistically independent data sets $A$
and $B$ should result in an acceptable $p$-value from the PG test.  To
illustrate an inconsistency in the data, it is however sufficient to
demonstrate that at least one way of dividing it leads to a poor
value. Here, we choose to split the data into disappearance data
encompassing the oscillation channels $\parenbar\nu_e \to
\parenbar\nu_e$ and $\parenbar\nu_\mu \to \parenbar\nu_\mu$, and
appearance data covering the $\parenbar\nu_\mu \to \parenbar\nu_e$
channel. Note that it is important to chose data sets independent of
their ``result''. For instance, dividing data into ``evidence'' and
``no-evidence'' samples would bias the PG test.

\begin{figure}
  \centering
  \includegraphics[width=.5\textwidth]{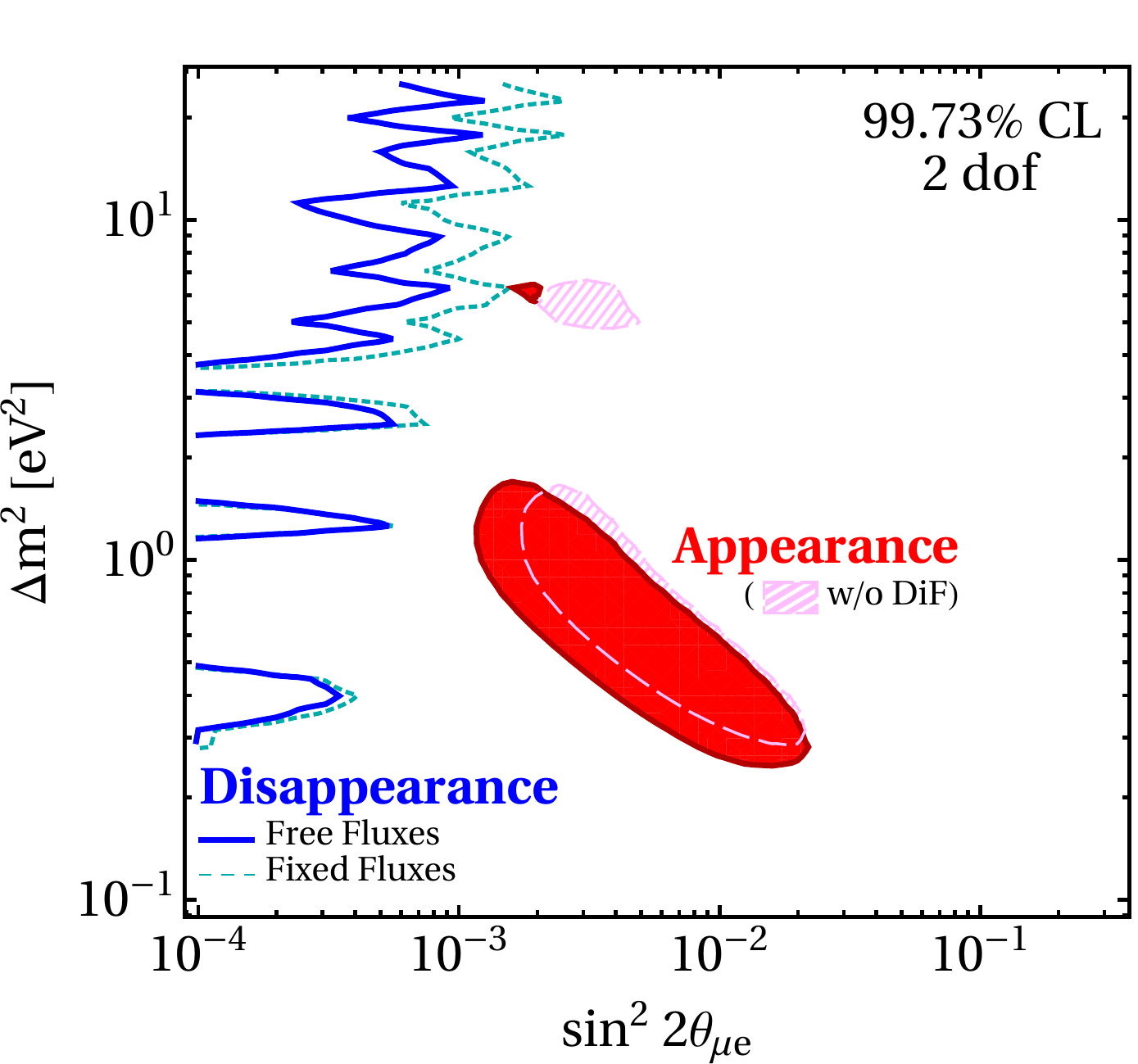}
  \caption{Appearance versus disappearance data in the plane spanned
    by the effective mixing angle $\sin^2 2\theta_{\mu
      e}\equiv4|U_{e4}U_{\mu4}|^2$ and the mass squared difference
    $\Delta m_{41}^2$. The blue curves show limits from the
    disappearance data sets using free reactor fluxes (solid) or fixed
    reactor fluxes (dashed), while the shaded contours are based on
    the appearance data sets using LSND DaR+DiF (red) and LSND DaR
    (pink hatched). All contours are at 99.73\%~CL for 2~dof.}
  \label{fig:app-vs-disapp}
\end{figure}

The tension between appearance and disappearance data is shown
graphically in \cref{fig:app-vs-disapp}. The figure illustrates the
lack of overlap between the parameter region favoured by appearance
data (driven by LSND and MiniBooNE) and the strong exclusion limits
from disappearance data. The tension persists independently of whether
reactor fluxes are fixed or kept free, and whether the LSND DaR or
DaR+DiF samples are used. The corresponding results from the PG test
are shown in the last two columns of \cref{tab:global-bf}. To evaluate
the $p$-value of the PG test statistic we use two degrees of freedom,
corresponding to the two parameters in common to appearance and
disappearance data, see \cref{tab:dofs} and the related discussion. We
observe that for none of the analyses given in the table, the
$p$-value for appearance and disappearance data being consistent
exceeds $10^{-5}$, with the ``best'' compatibility of $p = 2.6\times
10^{-6}$ emerging for fixed reactor fluxes and using LSND DaR+DiF
data.  We conclude that the appearance/disappearance tension excludes
a sterile neutrino oscillation explanation of the $\parenbar\nu_\mu
\to \parenbar\nu_e$ anomalies at the $4.7\sigma$ level.

Note that the parameter goodness-of-fit for the analysis using free reactor
fluxes is worse than the one for fixed reactor fluxes. The reason can be
understood from the $\chi^2$ numbers given in \cref{tab:global-bf}. We see that
the $\chi^2_\text{min}$ of $\parenbar\nu_e$ disappearance decreases by more
(9.9 units) than the global best fit point (7 or 6 units for DaR or DaR+DiF,
respectively), when leaving reactor fluxes free. Therefore, reactor data alone
benefits more from free fluxes than the appearance/disappearance tension, which
increases the $\chi^2$ penalty to pay for the combination in the case of free
fluxes.

\begin{table}
  \centering
  \begin{ruledtabular}
  \begin{tabular}{llccccccc}
    & Analysis         & $\chi^2_\text{min,global}$ & $\chi^2_\text{min,app}$ & $\Delta\chi^2_\text{app}$
                       & $\chi^2_\text{min,disapp}$ & $\Delta\chi^2_\text{disapp}$
                       & $\chi^2_\text{PG}$/dof & PG \\
    \hline          
    & Global          & 1120.9 & 79.1 & 11.9 & 1012.2 & 17.7 & 29.6/2 & $3.71 \times 10^{-7}$ \\
    \hline
    \multicolumn{8}{l}{\bf Removing anomalous data sets} \\ 
    & w/o LSND         & 1099.2 &  86.8 & 12.8 & 1012.2 & 0.1 & 12.9/2 & $1.6\times 10^{-3}$ \\
    & w/o MiniBooNE    & 1012.2 &  40.7 &  8.3 &  947.2 & 16.1 & 24.4/2 & $5.2\times 10^{-6}$ \\
    & w/o reactors     &  925.1 &  79.1 & 12.2 &  833.8 & 8.1  & 20.3/2 & $3.8\times 10^{-5}$ \\
    & w/o gallium      & 1116.0 &  79.1 & 13.8 & 1003.1 & 20.1 & 33.9/2 & $4.4\times 10^{-8}$ \\
    \hline
    \multicolumn{8}{l}{\bf Removing constraints} \\
    & w/o IceCube      &  920.8  & 79.1 & 11.9 & 812.4 & 17.5 & 29.4/2 & $4.2\times 10^{-7}$ \\
    & w/o MINOS(+)     &  1052.1 & 79.1 & 15.6 & 948.6 & 8.94 & 24.5/2 & $4.7\times 10^{-6}$ \\
    & w/o MB disapp    &  1054.9 & 79.1 & 14.7 & 947.2 & 13.9 & 28.7/2 & $6.0\times 10^{-7}$ \\
    & w/o CDHS         &  1104.8 & 79.1 & 11.9 & 997.5 & 16.3 & 28.2/2 & $7.5\times 10^{-7}$ \\
    \hline
    \multicolumn{8}{l}{\bf Removing classes of data} \\
    & $\protect\parenbar\nu_e$   dis vs app & 628.6 & 79.1 & 0.8 & 542.9 & 5.8 & 6.6/2 & $3.6\times 10^{-2}$ \\
    & $\protect\parenbar\nu_\mu$ dis vs app & 564.7 & 79.1 & 12.0 & 468.9 & 4.7 & 16.7/2 & $2.3\times 10^{-4}$ \\
    & $\protect\parenbar\nu_\mu$ dis + solar vs app & 884.4 & 79.1 & 13.9 & 781.7 & 9.7 & 23.6/2 & $7.4\times 10^{-6}$ \\
  \end{tabular}
  \end{ruledtabular}
  \caption{Results of the parameter goodness-of-fit (PG) test
    \cite{Maltoni:2003cu} comparing appearance to disappearance
    data. In this table we use the reactor flux-free analysis and LSND
    DaR+DiF data; therefore we do not quote dof for the $\chi^2$
    values. The first row corresponds to the global fit, while the
    other row show the impact of removing individual experiments or sets of experiments from
    the fit. In columns 2--8, we list the $\chi^2$ at the global best fit point
    ($\chi^2_\text{min,global}$), the $\chi^2$ at the
    appearance best fit ($\chi^2_\text{min,app}$), the difference
    in $\chi^2_\text{app}$ between the appearance best fit point and
    the global best fit point ($\Delta\chi^2_\text{app}$), the $\chi^2$
     at the disappearance best fit
    ($\chi^2_\text{min,disapp}$), the difference in
    $\chi^2_\text{disapp}$ between the disappearance best fit point
    and the global best fit point ($\Delta\chi^2_\text{disapp}$), the
    $\chi^2$ per dof for the PG test ($\chi^2_\text{PG}$/dof, computed
    according to \cref{eq:pg-chi2}), and the resulting $p$-value given
    by \cref{eq:p-PG}.}
  \label{tab:pg}
\end{table}

In \cref{tab:pg} we investigate the robustness of the
appearance/disappearance tension.  We show how the PG would improve if
individual experiments or classes of experiments were removed from the
fit. We stress that we are not aware of any strong reason to discard
data from particular experiments. The sole purpose of this exercise is
to demonstrate the impact of individual data sets and establish the
robustness of our conclusion.

The first row in \cref{tab:pg} corresponds to the global analysis
using free reactor fluxes and LSND DaR+DiF data, which is the combination of
data we use throughout this table. The remaining part of the table
shows that very strong tension remains even after removing any individual
experiment. In particular, the PG remains below $\approx 5\times
10^{-6}$ when any of the $\parenbar\nu_\mu$ disappearance data sets
are removed, so it does not rely on the particular treatment of any of
those experiments.  Even when \emph{all} reactor data are removed, the PG
remains very small ($3.8\times 10^{-5}$).

The only significant improvement is obtained when removing LSND. The
still somewhat low PG of $0.16\%$ is a manifestation of the tension
between the MiniBooNE excess and the disappearance data. But it is
clear that the very strong appearance/disappearance tension is driven
by LSND.  Note also that this remains true when MiniBooNE is removed,
and therefore the result does not depend on the low-energy excess in
MiniBooNE.

The only way to reconcile LSND would be to discard $\parenbar\nu_\mu$
disappearance data altogether. Note that even if we remove all
$\parenbar\nu_e$ disappearance data, the PG remains low, at $2.4\times
10^{-4}$. The reason is the non-trivial constraint on $|U_{e4}|$ from
the data sample we call $\parenbar\nu_\mu$ disappearance (defined in
\cref{tab:nu-mu-disapp}), see \cref{fig:nu-e}. Remarkably, just using
$\parenbar\nu_\mu$ disappearance plus solar neutrinos pushes the PG
already to $7.4\times 10^{-6}$. This demonstrates once again that our
conclusion is independent of reactor neutrino data.

We observe from \cref{tab:pg} that the PG gets nearly an order
of magnitude worse when removing the gallium data. The reason is the
slight tension between gallium and reactor data discussed in
\cref{sec:global_nu-e}. If gallium is removed, the $\parenbar\nu_e$
disappearance fit alone improves, and therefore the tension with
appearance data increases.

Finally, we have also performed a slightly different PG test, by
dividing the data into $\nu_\mu$ disappearance versus the combined
$\nu_e$ appearance and $\nu_e$ disappearance data. This corresponds to
the samples compared in \cref{fig:nu-mu}. Using LSND DaR+DiF data and
free reactor fluxes we obtain a $\chi^2_{\rm PG} = 23.4$. According to
\cref{tab:dofs}, the common parameters in those two data sets are
$\Delta m^2_{31}, \Delta m^2_{41},
\theta_{14},\theta_{24},\theta_{34}$. Therefore, $\chi^2_{\rm PG}$ has
to be evaluated for 5~dof, leading to a $p$-value of $2.8\times
10^{-4}$.

\section{Discussion and Conclusions}
\label{sec:discussion}

We have presented an updated global analysis of neutrino oscillation
data within a $3+1$ sterile neutrino mass scheme.  We have obtained
two main results, which can be summarized as follows:
\begin{enumerate}
\item Reactor neutrino data show a $\gtrsim 3\sigma$ preference for sterile
  neutrino oscillations with $\Delta m^2_{41} \approx 1.3$~eV$^2$ and $|U_{e4}|
  \approx 0.1$. This is driven by recent data from DANSS and NEOS and is based
  only on the relative comparisons of measured energy spectra and is therefore
  independent of predictions for the reactor neutrino fluxes and spectra.
  If flux predictions are taken into account, the
  preference for sterile neutrino oscillations in global $\parenbar\nu_e$
  disappearance data increases to $3.8\sigma$.
  
\item Constraints on $\parenbar\nu_\mu$ disappearance have become
  exceedingly strong, due to recent data from MINOS/MINOS+ and
  IceCube. This leads to very strong tension between the anomalies in the
  appearance sector (LSND and MiniBooNE) and disappearance data. We
  find that appearance and disappearance data are incompatible, with a
  parameter goodness-of-fit test yielding a $p$-value of less than
  $2.6\times 10^{-6}$. This result does not
  rely on any single experiment in the $\parenbar\nu_\mu$ sector and
  is robust with respect to theoretical predictions of reactor fluxes;
  the $p$-value remains at
  $3.8 \times 10^{-5}$ even if all reactor data are removed. The
  tension is dominated by LSND; the MiniBooNE anomaly plays a
  subleading role.
\end{enumerate}
Our results rule out the sterile neutrino oscillation hypothesis as
an explanation of the LSND and MiniBooNE anomalies, but it remains
a viable option for the reactor and gallium anomalies.

Some comments are in order. Our conclusion in item~1 above is largely
based on preliminary data from DANSS presented at
conferences~\cite{danss-moriond17,danss-solvay17}.  Our results are in
agreement with another recent analysis done outside the DANSS
collaboration~\cite{Gariazzo:2018mwd}. However, those results will
need to be supported by an official publication by the collaboration.

Throughout this work we have restricted ourselves to the $3+1$ scenario, adding just
one mass state at the eV scale.  However, we expect that the tension between
appearance and disappearance data cannot be resolved by adding more sterile
neutrinos. This has been quantitatively investigated previously,
e.g.~\cite{Maltoni:2007zf, Kopp:2013vaa}. There, it had been shown that adding
more neutrinos does not relax the tension. The reason is that the quadratic
suppression of the $\nu_\mu\to\nu_e$ oscillation amplitudes by constraints on
the elements $|U_{ei}|$ and $|U_{\mu i}|$ ($i \ge 4$) from disappearance data
remains equally true in scenarios with more than one eV-scale mass states.
Therefore we expect that our conclusion concerning the sterile neutrino
explanation of appearance anomalies remains qualitatively true also for more
sterile neutrinos.

Finally, we remind the reader that a completely orthogonal set of constraints
on eV-scale sterile neutrinos comes from cosmology.  The standard picture
is that active neutrinos evolve into a superposition of active and sterile
states at temperatures $\gtrsim \text{MeV}$. Hard, flavour-sensitive collisions
mediated by $W$ and $Z$ bosons collapse these superpositions into purely active
or purely sterile states, with the relative probability given by the
active--sterile mixing angles. After a large number of collisions, active and
sterile neutrinos come into thermal equilibrium.  Because of this, the vanilla
$3+1$ model appears to be strongly disfavoured by constraints on the number of
relativistic species $N_\text{eff}$ at the time of Big Bang Nucleosynthesis
(BBN) \cite{Cyburt:2015mya} and during the recombination epoch
\cite{Ade:2015xua}.  Moreover, constraints on the sum of neutrino masses, $\sum
m_\nu$ from Cosmic Microwave Background and structure formation data disfavour
extra neutrino species with masses $\gtrsim 0.3$~eV~\cite{Ade:2015xua}.
However, these constraints are model-dependent, and in non-minimal scenarios
they can be weakened or absent.  A full review of such scenarios is well beyond
the scope of this work, therefore we only mention a few exemplary ones: in
particular, mechanisms discussed in the literature include new interactions in
the sterile sector \cite{Dasgupta:2013zpn, Hannestad:2013ana,
Chu:inprogress, Cherry:2016jol}, an extremely low reheating temperature
\cite{Yaguna:2007wi}, large neutrino--anti-neutrino asymmetries
\cite{Saviano:2013ktj}, late entropy production \cite{Ho:2012br}, and the
presence of matter and antimatter domains during BBN \cite{Giovannini:2002qw}.
It is also worth noting that the prevailing tension between local and
cosmological determinations of the Hubble constant would be relaxed if
$N_\text{eff}$ is somewhat larger than in the SM~\cite{Bernal:2016gxb}.

\bigskip

\emph{Note added:} While we were finalizing this work, the STEREO
collaboration announced first results from their search for
short-baseline neutrino oscillations at the ILL reactor in
Grenoble~\cite{Lhuillier:Moriond2018}.  At the moment, we do not
expect these new exclusion limits to have a significant impact on the
preferred region for the combined reactor data and global
$\parenbar\nu_e$ disappearance data yet.

\section*{Acknowledgements}

We are grateful to Carlos Arg\"uelles, Janet Conrad, and William Louis
for useful discussions. We thank Gavin Davies for help in implementing
the NO$\nu$A results.  The work of JK and MD has been supported by the
German Research Foundation (DFG) under Grant
Nos.\ \mbox{KO~4820/1--1}, FOR~2239, EXC-1098 (PRISMA) and by the
European Research Council (ERC) under the European Union's Horizon
2020 research and innovation programme (grant agreement No.\ 637506,
``$\nu$Directions''). The work of MM and IMS has been supported by the
Spanish government through MINECO/FEDER-UE grants FPA2015-65929-P and
FPA2016-78645-P, as well as the ``Severo Ochoa'' program grant
SEV-2016-0597 of IFT. This project has received support from the
European Union's Horizon 2020 research and innovation programme under
the Marie Sk\l{}odowska-Curie grant agreements No 690575
(InvisiblesPlus) and No 674896 (Elusives).  This manuscript has been
authored by Fermi Research Alliance, LLC under Contract
No.\ DE-AC02-07CH11359 with the U.S.\ Department of Energy, Office of
Science, Office of High Energy Physics.

\appendix
\section{The parameter goodness-of-fit test}
\label{sec:PG}

In this appendix we briefly review the parameter goodness-of-fit (PG)
test~\cite{Maltoni:2003cu}, which measures the compatibility of
sub-sets of a data set. Let us subdivide the global data into two
statistically independent sets $A$ and $B$. Let
$\chi^2_{\text{min},A}$ and $\chi^2_{\text{min},B}$ be the minimum
$\chi^2$ values obtained from individual fits to the two data sets,
and let $\chi^2_\text{min,global}$ be the $\chi^2$ at the global best
fit point obtained from a combined fit to all the data.  The quantity
\begin{align}
  \chi^2_\text{PG} \equiv \chi^2_\text{min,global} - \chi^2_{\text{min},A} - \chi^2_{\text{min},B}
                   =      \Delta\chi^2_A + \Delta\chi^2_B
  \label{eq:pg-chi2}
\end{align}
measures by how much the fit worsens when the two data sets are
combined.  This can be seen from the second equality in
\cref{eq:pg-chi2}, in which we have defined, for each subset of the
data, the $\chi^2$ difference $\Delta\chi^2_{A,B}$ between the
individual best fit point and the global best fit point.  If
$\chi^2_{A}$ and $\chi^2_{B}$ depend on $P_A$
and $P_B$ parameters, respectively, and $P$ is the total number
of parameters of the model ($P_A,P_B \le P$),
then one can show~\cite{Maltoni:2003cu} that $\chi^2_\text{PG}$
follows a $\chi^2$ distribution with
\begin{align}
  N_\text{PG} \equiv P_A + P_B - P
  \label{eq:N-pg}
\end{align}
degrees of freedom.\footnote{$N_\text{PG}$ counts the number of
  ``joint'' parameters of the data sets $A$ and $B$.  As an example,
  if $A$ and $B$ depend on exactly the same $P$ parameters, then $P_A
  = P_B = N_\text{PG} = P$.}  We can thus compute a $p$-value
measuring the compatibility of the data sets $A$ and $B$ according to
\begin{align}
  p = \int_{\chi^2_\text{PG}}^\infty \! dx \, f_{\chi^2}(x; N_\text{PG}) \,,
  \label{eq:p-PG}
\end{align}
where $f_{\chi^2}(x; N_\text{PG})$ is the probability density function of the $\chi^2$
distribution with $N_\text{PG}$ degrees of freedom.

\section{Details of the IceCube Fit}
\label{sec:icecube-fit}

The event numbers measured by the IceCube detector have been provided
in a grid with 210 bins~\cite{Jones:2015, TheIceCube:2016oqi}, which depends
on the reconstructed muon energy $E_\mu$ (logarithmically spaced in 10
bins ranging from 400~GeV to 20~TeV) and the reconstructed muon
direction (linearly spaced in 21 bins from $\cos\theta = -1.02$ to
$\cos\theta = 0.24$). We make the assumption that the reconstructed muon 
direction is the same as the direction of the initial neutrino. The predicted
number of events in bin number $(ij)$ (where $i$ indexes $\cos\theta$
and $j$ indexes $E_\mu$) is computed according to
\begin{align}
  N^{d,f}_{ij} &= \int \! dE_\nu \Big[
     \phi^{\text{atm},f}_{+}(E_\nu,\theta^i,N_0,\gamma,R_{\pi/K}) \,
       \bar{P}_{\mu\mu}^+(E_\nu,\theta^i) \,
       A^d_\text{eff,+}(E_\nu,E_\mu^j,\theta^i) \nonumber \\ 
    &\quad + 
      R_{\pm}\,\phi^{\text{atm},f}_{-}(E_\nu,\theta^i,N_0,\gamma,R_{\pi/K}) \,
      \bar {P}_{\mu\mu}^-(E_\nu,\theta^i) \,
      A^d_{\text{eff},-}(E_\nu,E_\mu^j,\theta^i) \Big]\,.
  \label{eq:N-IC}
\end{align}
Here, $\phi^{\text{atm},f}_\pm(E_\nu,\theta^i,N_0,\gamma,R_{\pi/K})$
is the atmospheric muon neutrino ($+$) or anti-neutrino ($-$) flux, which depends on
the true neutrino energy $E_\nu$, the neutrino direction $\theta^i$, and on the
nuisance parameters $N_0$, $\gamma$, and $R_{\pi/K}$ discussed below. It also
depends on the theoretical flux model, indicated by the subscript $f$.
The effective area $A^d_{\text{eff},\pm}(E_\nu,E_\mu^j, \theta^i)$ in
\cref{eq:N-IC} encodes the detector response to a $\nu_\mu$ ($+$) or
$\bar\nu_\mu$ ($-$) with energy $E_\nu$ and direction $\theta^i$.  The IceCube
collaboration provides $A^d_{\text{eff},\pm}(E_\nu,E_\mu^j, \theta^i)$
in the form of a three-dimensional array in $E_\mu$, $\cos\theta$ (same binning
as for the data), and $E_\nu$ (200~bins logarithmically spaced between 200~GeV
and 1~PeV) \cite{TheIceCube:2016oqi}. Separate arrays are provided for different
assumptions on the Digital Optical Module (DOM) efficiency, indicated by the
superscript $d$. 

The muon neutrino and anti-neutrino survival probability $\bar{P}_{\mu\mu}^\pm$
is computed using GLoBES~\cite{Huber:2004ka, Huber:2007ji}, including a
low-pass filter to suppress fast oscillation and to account for the limited
energy resolution of the detector. For the production height of the neutrinos
we interpolate linearly between 28~km for horizontal neutrinos and 18~km for
vertical neutrinos~\cite{Honda:2015fha}. To model the attenuation of the
neutrino flux due to absorption in the Earth, we multiply the oscillation
probability by an exponential damping factor given by 
\begin{align}
  e^{-X(\theta) \, \sigma^\pm(E) (1-P^{\pm}_{\mu\mu})},
   \label{eq:attenuation}
\end{align}
where $X(\theta)$ is the column density along the neutrino trajectory and
$\sigma^\pm(E)$ the inclusive absorption cross-section for neutrinos and
antineutrinos, respectively. The factor $(1-P^{\pm}_{\mu\mu}(E,L))$ accounts
for the fact that only the active flavors interact with matter. This formula
holds exactly only for an oscillation probability independent of the length of
the trajectory. We make the assumption that in much of the parameter space the
oscillations are either averaged out, or the oscillation length is so long that
the probability is approximately constant along the trajectory. We have checked
that our results do not depend significantly on this assumption.\\
 
In the published IceCube fit~\cite{TheIceCube:2016oqi}, systematic
uncertainties are included either as discrete or as continuous nuisance
parameters.  The only discrete nuisance parameter in our analysis is the
theoretical flux model.  We found that out of the seven
flux models considered by the IceCube collaboration, only two contribute
significantly, namely the ones tagged ``PolyGonato QGSJET--II--04'' and
``Honda-Gaisser''. We therefore restrict our analysis to these two discrete
models. Hence the index $f$ in \cref{eq:N-IC} runs from 1
to 2.

The continuous nuisance parameters can be divided into two classes: those
related to the neutrino flux, and those related to the detector response and
the optical properties of the ice.  In our analysis we use the following
atmospheric neutrino flux uncertainties:
\begin{itemize}
  \item the normalization $N_0$. Formally we assume a large
    uncertainty of 40\% on the normalization, but results are very
    similar for completely free normalization. Therefore we consider
    $N_0$ to be effectively unconstrained.
  
  \item the tilt of the energy spectrum, which is parameterized by including
    a factor $(E/E_0)^{\gamma}$, with a 5\% error on the power law index $\gamma$
    and a central value of $\gamma = 0$;

  \item the ratio between the pion and the kaon decay contributions to
    the flux, $R_{\pi/K}$, with an error of 10\%;

  \item the ratio between the neutrino and the anti-neutrino fluxes,
    $R_{\pm}$, with an error of 2.5\%.
\end{itemize}
Out of the uncertainties associated with the detector response and the ice
properties, we only include the uncertainty on the DOM efficiency. As 
stated above, the tabulated effective area is provided for four different 
models for the DOM efficiency. We interpolate linearly between the 
per-bin-prediction for each DOM model and allow the minimizer to choose the
optimal superposition of DOM models. Concerning the ice 
properties, we restricted ourselves to the nominal model because effective areas
for each DOM efficiencies are only provided for the nominal ice 
model.

For each point in the parameter space a $\chi^2$ value is calculated
from the theoretical predictions and the experimental values by means
of a log-likelihood function.

\bigskip

We have cross-check our IceCube fit with a second version of the
analysis, which was developed completely independently. This analysis
is not using the GLoBES software but is based on a dedicated
probability code and it uses a partially different approach to
systematics.  The most noteworthy difference is the treatment of the
discrete systematics.  In our second implementation we restrict
ourselves to only one flux model, the ``Honda-Gaisser-model''. Several
other discrete systematics associated with the detector response are
treated as continuous quantities, and their effects on the number of
events are assumed to be linear. In detail, in our second
implementation we use:
\begin{itemize}
  \item the DOM efficiency, where as nominal value we have used the
    table corresponding to 99\% efficiency, and as $1\sigma$ deviation
    we have used the table corresponding to 95\% efficiency;

  \item photon scattering in the ice, where the $1\sigma$ deviation is
    defined from the table corresponding to a 10\% increase with respect
    to the nominal response;

  \item photon absorption in the ice, where the $1\sigma$ deviation is
    defined as a 10\% increase in the absorption rate with respect to
    the nominal response;

  \item the azimuthal anisotropy in the scattering length due to the dust
    grain shear; here the $1\sigma$ deviation is obtained from the data
    set denoted `SPICELEA ice model';

  \item the optical properties of the ice column surrounding each
    string, where the $1\sigma$ deviation is obtained from the data set
    labelled `SPICEMIE ice model'. This data set does not include hole ice 
    effects.
\end{itemize}
Furthermore, in our second implementation, we average the oscillation 
probability over the altitude of the neutrino production point. The averaged 
probability is given by
\begin{align}
  \ev{P_{\mu\mu}^\pm(E_\nu,\theta)}
  = \int dh \,
  P_{\mu\mu}^\pm(E_\nu,\cos\theta, h) \,
  \kappa^{\pm}(E_\nu,\cos\theta,h) \,,
  \label{eq:Pmm}
\end{align}
where $P_{\mu\mu}^\pm(E_\nu,\cos\theta, h)$ is the unaveraged oscillation
probability for a neutrino produced at altitude $h$ and
$\kappa^\pm(E_\nu,\cos\theta,h)$ is the distribution of production
altitudes, normalized to one~\cite{Honda:2015fha}. 

We find good agreement between our two implementations, and between each of our
implementations and the official IceCube results~\cite{TheIceCube:2016oqi}. We
therefore conclude that our IceCube analysis is robust.

\bibliographystyle{JHEP}
\bibliography{refs}

\providecommand{\href}[2]{#2}\begingroup\raggedright\begin{thebibliography}{100}

\bibitem{Aguilar:2001ty}
{\bf LSND} {\bf Collaboration}, A.~Aguilar {\em et~al.}, {\it {Evidence for
  neutrino oscillations from the observation of $\bar{\nu}_e$ appearance in a
  $\bar{\nu}_\mu$ beam}},  {\em Phys. Rev.} {\bf D64} (2001) 112007,
  [\href{http://www.arxiv.org/abs/hep-ex/0104049}{{\tt hep-ex/0104049}}].

\bibitem{Aguilar-Arevalo:2013pmq}
{\bf MiniBooNE} {\bf Collaboration}, A.~Aguilar-Arevalo {\em et~al.}, {\it
  {Improved Search for $\bar \nu_\mu \rightarrow \bar \nu_e$ Oscillations in
  the MiniBooNE Experiment}},  {\em Phys.Rev.Lett.} {\bf 110} (2013) 161801,
  [\href{http://www.arxiv.org/abs/1207.4809}{{\tt 1207.4809}}].

\bibitem{Mueller:2011nm}
T.~Mueller, D.~Lhuillier, M.~Fallot, A.~Letourneau, S.~Cormon, {\em et~al.},
  {\it {Improved Predictions of Reactor Antineutrino Spectra}},  {\em
  Phys.Rev.} {\bf C83} (2011) 054615,
  [\href{http://www.arxiv.org/abs/1101.2663}{{\tt 1101.2663}}].

\bibitem{Huber:2011wv}
P.~Huber, {\it {On the determination of anti-neutrino spectra from nuclear
  reactors}},  {\em Phys.Rev.} {\bf C84} (2011) 024617,
  [\href{http://www.arxiv.org/abs/1106.0687}{{\tt 1106.0687}}].

\bibitem{Mention:2011rk}
G.~Mention, M.~Fechner, T.~Lasserre, T.~Mueller, D.~Lhuillier, {\em et~al.},
  {\it {The Reactor Antineutrino Anomaly}},  {\em Phys.Rev.} {\bf D83} (2011)
  073006, [\href{http://www.arxiv.org/abs/1101.2755}{{\tt 1101.2755}}].

\bibitem{Hayes:2013wra}
A.~C. Hayes, J.~L. Friar, G.~T. Garvey, G.~Jungman, and G.~Jonkmans, {\it
  {Systematic Uncertainties in the Analysis of the Reactor Neutrino Anomaly}},
  {\em Phys. Rev. Lett.} {\bf 112} (2014) 202501,
  [\href{http://www.arxiv.org/abs/1309.4146}{{\tt 1309.4146}}].

\bibitem{Fang:2015cma}
D.-L. Fang and B.~A. Brown, {\it {Effect of first forbidden decays on the shape
  of neutrino spectra}},  {\em Phys. Rev.} {\bf C91} (2015), no.~2 025503,
  [\href{http://www.arxiv.org/abs/1502.02246}{{\tt 1502.02246}}]. [Erratum:
  Phys. Rev.C93,no.4,049903(2016)].

\bibitem{Hayes:2016qnu}
A.~C. Hayes and P.~Vogel, {\it {Reactor Neutrino Spectra}},  {\em Ann. Rev.
  Nucl. Part. Sci.} {\bf 66} (2016) 219--244,
  [\href{http://www.arxiv.org/abs/1605.02047}{{\tt 1605.02047}}].

\bibitem{Acero:2007su}
M.~A. Acero, C.~Giunti, and M.~Laveder, {\it {Limits on $\nu_e$ and $\bar\nu_e$
  disappearance from Gallium and reactor experiments}},  {\em Phys.Rev.} {\bf
  D78} (2008) 073009, [\href{http://www.arxiv.org/abs/0711.4222}{{\tt
  0711.4222}}].

\bibitem{Giunti:2010zu}
C.~Giunti and M.~Laveder, {\it {Statistical Significance of the Gallium
  Anomaly}},  {\em Phys.Rev.} {\bf C83} (2011) 065504,
  [\href{http://www.arxiv.org/abs/1006.3244}{{\tt 1006.3244}}].

\bibitem{Kopp:2011qd}
J.~Kopp, M.~Maltoni, and T.~Schwetz, {\it {Are there sterile neutrinos at the
  eV scale?}},  {\em Phys.Rev.Lett.} {\bf 107} (2011) 091801,
  [\href{http://www.arxiv.org/abs/1103.4570}{{\tt 1103.4570}}].

\bibitem{Conrad:2012qt}
J.~Conrad, C.~Ignarra, G.~Karagiorgi, M.~Shaevitz, and J.~Spitz, {\it {Sterile
  Neutrino Fits to Short Baseline Neutrino Oscillation Measurements}},  {\em
  Adv.High Energy Phys.} {\bf 2013} (2013) 163897,
  [\href{http://www.arxiv.org/abs/1207.4765}{{\tt 1207.4765}}].

\bibitem{Archidiacono:2013xxa}
M.~Archidiacono, N.~Fornengo, C.~Giunti, S.~Hannestad, and A.~Melchiorri, {\it
  {Sterile neutrinos: Cosmology versus short-baseline experiments}},  {\em
  Phys. Rev.} {\bf D87} (2013), no.~12 125034,
  [\href{http://www.arxiv.org/abs/1302.6720}{{\tt 1302.6720}}].

\bibitem{Kopp:2013vaa}
J.~Kopp, P.~A.~N. Machado, M.~Maltoni, and T.~Schwetz, {\it {Sterile Neutrino
  Oscillations: The Global Picture}},  {\em JHEP} {\bf 1305} (2013) 050,
  [\href{http://www.arxiv.org/abs/1303.3011}{{\tt 1303.3011}}].

\bibitem{Mirizzi:2013kva}
A.~Mirizzi, G.~Mangano, N.~Saviano, E.~Borriello, C.~Giunti, {\em et~al.}, {\it
  {The strongest bounds on active-sterile neutrino mixing after Planck data}},
  \href{http://www.arxiv.org/abs/1303.5368}{{\tt 1303.5368}}.

\bibitem{Giunti:2013aea}
C.~Giunti, M.~Laveder, Y.~Li, and H.~Long, {\it {Pragmatic View of
  Short-Baseline Neutrino Oscillations}},  {\em Phys.Rev.} {\bf D88} (2013)
  073008, [\href{http://www.arxiv.org/abs/1308.5288}{{\tt 1308.5288}}].

\bibitem{Gariazzo:2013gua}
S.~Gariazzo, C.~Giunti, and M.~Laveder, {\it {Light Sterile Neutrinos in
  Cosmology and Short-Baseline Oscillation Experiments}},  {\em JHEP} {\bf
  1311} (2013) 211, [\href{http://www.arxiv.org/abs/1309.3192}{{\tt
  1309.3192}}].

\bibitem{Collin:2016rao}
G.~H. Collin, C.~A. Argüelles, J.~M. Conrad, and M.~H. Shaevitz, {\it {Sterile
  Neutrino Fits to Short Baseline Data}},  {\em Nucl. Phys.} {\bf B908} (2016)
  354--365, [\href{http://www.arxiv.org/abs/1602.00671}{{\tt 1602.00671}}].

\bibitem{Gariazzo:2017fdh}
S.~Gariazzo, C.~Giunti, M.~Laveder, and Y.~F. Li, {\it {Updated Global 3+1
  Analysis of Short-Baseline Neutrino Oscillations}},  {\em JHEP} {\bf 06}
  (2017) 135, [\href{http://www.arxiv.org/abs/1703.00860}{{\tt 1703.00860}}].

\bibitem{Giunti:2017yid}
C.~Giunti, X.~P. Ji, M.~Laveder, Y.~F. Li, and B.~R. Littlejohn, {\it {Reactor
  Fuel Fraction Information on the Antineutrino Anomaly}},  {\em JHEP} {\bf 10}
  (2017) 143, [\href{http://www.arxiv.org/abs/1708.01133}{{\tt 1708.01133}}].

\bibitem{Dentler:2017tkw}
M.~Dentler, A.~Hernández-Cabezudo, J.~Kopp, M.~Maltoni, and T.~Schwetz, {\it
  {Sterile Neutrinos Or Flux Uncertainties? — Status of the Reactor
  Anti-Neutrino Anomaly}},  {\em JHEP} {\bf 11} (2017) 099,
  [\href{http://www.arxiv.org/abs/1709.04294}{{\tt 1709.04294}}].

\bibitem{An:2016luf}
{\bf Daya Bay} {\bf Collaboration}, F.~P. An {\em et~al.}, {\it {Improved
  Search for a Light Sterile Neutrino with the Full Configuration of the Daya
  Bay Experiment}},  {\em Phys. Rev. Lett.} {\bf 117} (2016), no.~15 151802,
  [\href{http://www.arxiv.org/abs/1607.01174}{{\tt 1607.01174}}].

\bibitem{Ko:2016owz}
Y.~Ko {\em et~al.}, {\it {Sterile Neutrino Search at the Neos Experiment}},
  {\em Phys. Rev. Lett.} {\bf 118} (2017), no.~12 121802,
  [\href{http://www.arxiv.org/abs/1610.05134}{{\tt 1610.05134}}].

\bibitem{Alekseev:2016llm}
I.~Alekseev {\em et~al.}, {\it {Danss: Detector of the Reactor Antineutrino
  Based on Solid Scintillator}},  {\em JINST} {\bf 11} (2016), no.~11 P11011,
  [\href{http://www.arxiv.org/abs/1606.02896}{{\tt 1606.02896}}].

\bibitem{danss-moriond17}
M.~Danilov, 2017.
\newblock talk given on behalf of the DANSS Collaboration at the 52nd
  Rencontres de Moriond EW 2017, La Thuile, Italy; {\tt
  https://indico.in2p3.fr/event/13763/}.

\bibitem{danss-solvay17}
M.~Danilov, 2017.
\newblock talk given on behalf of the DANSS Collaboration at the Solvay
  Workshop 'Beyond the Standard model with Neutrinos and Nuclear Physics', 29
  Nov.--1 Dec.\ 2017, Brussels, Belgium.

\bibitem{Seo:2016uom}
{\bf RENO} {\bf Collaboration}, S.~H. Seo {\em et~al.}, {\it {Spectral
  Measurement of the Electron Antineutrino Oscillation Amplitude and Frequency
  using 500 Live Days of RENO Data}},
  \href{http://www.arxiv.org/abs/1610.04326}{{\tt 1610.04326}}.

\bibitem{Abe:2014bwa}
{\bf Double Chooz} {\bf Collaboration}, Y.~Abe {\em et~al.}, {\it {Improved
  Measurements of the Neutrino Mixing Angle $\theta_{13}$ with the Double Chooz
  Detector}},  {\em JHEP} {\bf 10} (2014) 086,
  [\href{http://www.arxiv.org/abs/1406.7763}{{\tt 1406.7763}}]. [Erratum:
  JHEP02,074(2015)].

\bibitem{An:2016srz}
{\bf Daya Bay} {\bf Collaboration}, F.~P. An {\em et~al.}, {\it {Improved
  Measurement of the Reactor Antineutrino Flux and Spectrum at Daya Bay}},
  {\em Chin. Phys.} {\bf C41} (2017), no.~1 013002,
  [\href{http://www.arxiv.org/abs/1607.05378}{{\tt 1607.05378}}].

\bibitem{Dwyer:2014eka}
D.~A. Dwyer and T.~J. Langford, {\it {Spectral Structure of Electron
  Antineutrinos from Nuclear Reactors}},  {\em Phys. Rev. Lett.} {\bf 114}
  (2015), no.~1 012502, [\href{http://www.arxiv.org/abs/1407.1281}{{\tt
  1407.1281}}].

\bibitem{Hayes:2015yka}
A.~C. Hayes, J.~L. Friar, G.~T. Garvey, D.~Ibeling, G.~Jungman, T.~Kawano, and
  R.~W. Mills, {\it {Possible Origins and Implications of the Shoulder in
  Reactor Neutrino Spectra}},  {\em Phys. Rev.} {\bf D92} (2015), no.~3 033015,
  [\href{http://www.arxiv.org/abs/1506.00583}{{\tt 1506.00583}}].

\bibitem{Novella:2015eaw}
P.~Novella, {\it {The Antineutrino Energy Structure in Reactor Experiments}},
  {\em Adv. High Energy Phys.} {\bf 2015} (2015) 364392,
  [\href{http://www.arxiv.org/abs/1512.03366}{{\tt 1512.03366}}].

\bibitem{Giunti:2016elf}
C.~Giunti, {\it {Precise Determination of the $^{235}$U Reactor Antineutrino
  Cross Section Per Fission}},  {\em Phys. Lett.} {\bf B764} (2017) 145--149,
  [\href{http://www.arxiv.org/abs/1608.04096}{{\tt 1608.04096}}].

\bibitem{Huber:2016xis}
P.~Huber, {\it {Neos Data and the Origin of the 5 Mev Bump in the Reactor
  Antineutrino Spectrum}},  {\em Phys. Rev. Lett.} {\bf 118} (2017), no.~4
  042502, [\href{http://www.arxiv.org/abs/1609.03910}{{\tt 1609.03910}}].

\bibitem{Mohanty:2017iyh}
A.~K. Mohanty, {\it {Possible origin of shoulder in the reactor antineutrino
  spectrum}},  \href{http://www.arxiv.org/abs/1711.02801}{{\tt 1711.02801}}.

\bibitem{Berryman:2018jxt}
J.~M. Berryman, V.~Brdar, and P.~Huber, {\it {Nuclear and Particle Conspiracy
  Solves Both Reactor Antineutrino Anomalies}},
  \href{http://www.arxiv.org/abs/1803.08506}{{\tt 1803.08506}}.

\bibitem{An:2017osx}
{\bf Daya Bay} {\bf Collaboration}, F.~P. An {\em et~al.}, {\it {Evolution of
  the Reactor Antineutrino Flux and Spectrum at Daya Bay}},  {\em Phys. Rev.
  Lett.} {\bf 118} (2017), no.~25 251801,
  [\href{http://www.arxiv.org/abs/1704.01082}{{\tt 1704.01082}}].

\bibitem{Huber:2015ouo}
P.~Huber and P.~Jaffke, {\it {Neutron capture and the antineutrino yield from
  nuclear reactors}},  {\em Phys. Rev. Lett.} {\bf 116} (2016), no.~12 122503,
  [\href{http://www.arxiv.org/abs/1510.08948}{{\tt 1510.08948}}].

\bibitem{Huber:nuplatformweek}
P.~Huber, ``Nuclear physics and the reactor anomaly.''
\newblock talk given at the CERN Neutrino Platform Week, Jan 29--Feb 2, 2018.
  slides availbale at
  \url{https://indico.cern.ch/event/645835/contributions/2777775/attachments/1593138/2522053/cern2018.pdf}.

\bibitem{Agafonova:2013xsk}
{\bf OPERA} {\bf Collaboration}, N.~Agafonova {\em et~al.}, {\it {Search for
  $\nu_\mu \rightarrow \nu_e$ oscillations with the OPERA experiment in the
  CNGS beam}},  {\em JHEP} {\bf 07} (2013) 004,
  [\href{http://www.arxiv.org/abs/1303.3953}{{\tt 1303.3953}}]. [Addendum:
  JHEP07,085(2013)].

\bibitem{Antonello:2012fu}
M.~Antonello, B.~Baibussinov, P.~Benetti, E.~Calligarich, N.~Canci, {\em
  et~al.}, {\it {Experimental search for the LSND anomaly with the ICARUS LAr
  TPC detector in the CNGS beam}},
  \href{http://www.arxiv.org/abs/1209.0122}{{\tt 1209.0122}}.

\bibitem{Farese:2014}
C.~Farese, {\it Results from icarus},  2014.
\newblock talk given at the Neutrino 2014 conference in Boston, slides
  available at
  \url{https://indico.fnal.gov/materialDisplay.py?contribId=265&sessionId=18&materialId=slides&confId=8022}.

\bibitem{Adamson:2017uda}
{\bf MINOS} {\bf Collaboration}, P.~Adamson {\em et~al.}, {\it {Search for
  Sterile Neutrinos in MINOS and MINOS+ Using a Two-Detector Fit}},  {\em
  Submitted to: Phys. Rev. Lett.} (2017)
  [\href{http://www.arxiv.org/abs/1710.06488}{{\tt 1710.06488}}].

\bibitem{Adamson:2017zcg}
{\bf NOvA} {\bf Collaboration}, P.~Adamson {\em et~al.}, {\it {Search for
  Active-Sterile Neutrino Mixing Using Neutral-Current Interactions in NOvA}},
  {\em Phys. Rev.} {\bf D96} (2017), no.~7 072006,
  [\href{http://www.arxiv.org/abs/1706.04592}{{\tt 1706.04592}}].

\bibitem{sksol:nakano2016}
Y.~Nakano, {\em {$^8$B solar neutrino spectrum measurement using
  Super-Kamiokande IV}}.
\newblock PhD thesis, Tokyo U., 2016-02.

\bibitem{Bellini:2014uqa}
{\bf Borexino} {\bf Collaboration}, G.~Bellini {\em et~al.}, {\it {Neutrinos
  from the primary proton–proton fusion process in the Sun}},  {\em Nature}
  {\bf 512} (2014), no.~7515 383--386.

\bibitem{Vinyoles:2016djt}
N.~Vinyoles, A.~M. Serenelli, F.~L. Villante, S.~Basu, J.~Bergström, M.~C.
  Gonzalez-Garcia, M.~Maltoni, C.~Peña-Garay, and N.~Song, {\it {A new
  Generation of Standard Solar Models}},  {\em Astrophys. J.} {\bf 835} (2017),
  no.~2 202, [\href{http://www.arxiv.org/abs/1611.09867}{{\tt 1611.09867}}].

\bibitem{wendell:2014dka}
{\bf Super-Kamiokande} {\bf Collaboration}, R.~Wendell, {\it {Atmospheric
  Results from Super-Kamiokande}},  {\em AIP Conf. Proc.} {\bf 1666} (2015)
  100001, [\href{http://www.arxiv.org/abs/1412.5234}{{\tt 1412.5234}}]. slides
  available at
  \url{https://indico.fnal.gov/event/8022/other-view?view=standard}.

\bibitem{Aartsen:2014yll}
{\bf IceCube} {\bf Collaboration}, M.~Aartsen {\em et~al.}, {\it {Determining
  neutrino oscillation parameters from atmospheric muon neutrino disappearance
  with three years of IceCube DeepCore data}},  {\em Phys. Rev.} {\bf D91}
  (2015), no.~7 072004, [\href{http://www.arxiv.org/abs/1410.7227}{{\tt
  1410.7227}}].

\bibitem{deepcore:2016}
{\bf IceCube} {\bf Collaboration}, J.~P. Yañez {\em et~al.}, ``{IceCube
  Oscillations: 3 years muon neutrino disappearance data}.''
  \href{http://icecube.wisc.edu/science/data/nu_osc}{\tt
  http://icecube.wisc.edu/science/data/nu\_osc}.

\bibitem{Honda:2015fha}
M.~Honda, M.~Sajjad~Athar, T.~Kajita, K.~Kasahara, and S.~Midorikawa, {\it
  {Atmospheric neutrino flux calculation using the NRLMSISE-00 atmospheric
  model}},  {\em Phys. Rev.} {\bf D92} (2015), no.~2 023004,
  [\href{http://www.arxiv.org/abs/1502.03916}{{\tt 1502.03916}}].

\bibitem{TheIceCube:2016oqi}
{\bf IceCube} {\bf Collaboration}, M.~G. Aartsen {\em et~al.}, {\it {Searches
  for Sterile Neutrinos with the Icecube Detector}},  {\em Phys. Rev. Lett.}
  {\bf 117} (2016), no.~7 071801,
  [\href{http://www.arxiv.org/abs/1605.01990}{{\tt 1605.01990}}]. Data
  accessible at
  \href{http://icecube.wisc.edu/science/data/IC86-sterile-neutrino}{\tt
  http://icecube.wisc.edu/science/data/IC86-sterile-neutrino}.

\bibitem{Jones:2015}
B.~J.~P. Jones, {\em Sterile neutrinos in cold climates}.
\newblock PhD thesis, Massachusetts Institute of Technology, 2015.
\newblock available from \url{http://hdl.handle.net/1721.1/101327}.

\bibitem{Arguelles:2015}
C.~A. {Arg\"{u}elles}, {\em New Physics with Atmospheric Neutrinos}.
\newblock PhD thesis, University of Wisconsin, Madison, 2015.
\newblock available from
  \url{https://docushare.icecube.wisc.edu/dsweb/Get/Document-75669/tesis.pdf}.

\bibitem{Wolfenstein:1977ue}
L.~Wolfenstein, {\it {Neutrino Oscillations in Matter}},  {\em Phys. Rev.} {\bf
  D17} (1978) 2369--2374.

\bibitem{Mikheev:1986gs}
S.~P. Mikheev and A.~{\relax Yu}. Smirnov, {\it {Resonance Amplification of
  Oscillations in Matter and Spectroscopy of Solar Neutrinos}},  {\em Sov. J.
  Nucl. Phys.} {\bf 42} (1985) 913--917. [Yad. Fiz.42,1441(1985)].

\bibitem{Nunokawa:2003ep}
H.~Nunokawa, O.~L.~G. Peres, and R.~Zukanovich~Funchal, {\it {Probing the Lsnd
  Mass Scale and Four Neutrino Scenarios with a Neutrino Telescope}},  {\em
  Phys. Lett.} {\bf B562} (2003) 279--290,
  [\href{http://www.arxiv.org/abs/hep-ph/0302039}{{\tt hep-ph/0302039}}].

\bibitem{Choubey:2007ji}
S.~Choubey, {\it {Signature of sterile species in atmospheric neutrino data at
  neutrino telescopes}},  {\em JHEP} {\bf 12} (2007) 014,
  [\href{http://www.arxiv.org/abs/0709.1937}{{\tt 0709.1937}}].

\bibitem{Kwon:1981ua}
H.~Kwon, F.~Boehm, A.~Hahn, H.~Henrikson, J.~Vuilleumier, {\em et~al.}, {\it
  {Search for neutrino oscillations at a fission reactor}},  {\em Phys.Rev.}
  {\bf D24} (1981) 1097--1111.

\bibitem{Zacek:1986cu}
{\bf CALTECH-SIN-TUM} {\bf Collaboration}, G.~Zacek {\em et~al.}, {\it
  {Neutrino Oscillation Experiments at the Gosgen Nuclear Power Reactor}},
  {\em Phys.Rev.} {\bf D34} (1986) 2621--2636.

\bibitem{Vidyakin:1987ue}
G.~Vidyakin, V.~Vyrodov, I.~Gurevich, Y.~Kozlov, V.~Martemyanov, {\em et~al.},
  {\it Detection of anti-neutrinos in the flux from two reactors},  {\em
  Sov.Phys.JETP} {\bf 66} (1987) 243--247.

\bibitem{Vidyakin:1994ut}
G.~S. Vidyakin {\em et~al.}, {\it {Limitations on the characteristics of
  neutrino oscillations}},  {\em JETP Lett.} {\bf 59} (1994) 390--393. [Pisma
  Zh. Eksp. Teor. Fiz.59,364(1994)].

\bibitem{Kozlov:1999cs}
{\relax Yu}.~V. Kozlov {\em et~al.}, {\it {Today and Future Neutrino
  Experiments at Krasnoyarsk Nuclear Reactor}},  {\em Nucl. Phys. Proc. Suppl.}
  {\bf 87} (2000) 514--516,
  [\href{http://www.arxiv.org/abs/hep-ex/9912046}{{\tt hep-ex/9912046}}].

\bibitem{Afonin:1988gx}
A.~Afonin, S.~Ketov, V.~Kopeikin, L.~Mikaelyan, M.~Skorokhvatov, {\em et~al.},
  {\it {A study of the reaction $\bar\nu_e + p \to e^+ + n$ on a nuclear
  reactor}},  {\em Sov.Phys.JETP} {\bf 67} (1988) 213--221.

\bibitem{Kuvshinnikov:1990ry}
A.~Kuvshinnikov, L.~Mikaelyan, S.~Nikolaev, M.~Skorokhvatov, and A.~Etenko,
  {\it {Measuring the $\bar\nu_e + p \to n + e^+$ cross-section and beta decay
  axial constant in a new experiment at Rovno NPP reactor. (In Russian)}},
  {\em JETP Lett.} {\bf 54} (1991) 253--257.

\bibitem{Declais:1994su}
Y.~Declais, J.~Favier, A.~Metref, H.~Pessard, B.~Achkar, {\em et~al.}, {\it
  {Search for neutrino oscillations at 15-meters, 40-meters, and 95-meters from
  a nuclear power reactor at Bugey}},  {\em Nucl.Phys.} {\bf B434} (1995)
  503--534.

\bibitem{Declais:1994ma}
Y.~Declais, H.~de~Kerret, B.~Lefievre, M.~Obolensky, A.~Etenko, {\em et~al.},
  {\it {Study of reactor anti-neutrino interaction with proton at Bugey nuclear
  power plant}},  {\em Phys.Lett.} {\bf B338} (1994) 383--389.

\bibitem{Greenwood:1996pb}
Z.~Greenwood, W.~Kropp, M.~Mandelkern, S.~Nakamura, E.~Pasierb-Love, {\em
  et~al.}, {\it {Results of a two position reactor neutrino oscillation
  experiment}},  {\em Phys.Rev.} {\bf D53} (1996) 6054--6064.

\bibitem{reno-EPS17}
H.~Seo, 2017.
\newblock talk given on behalf of the RENO Collaboration at the EPS conference
  on High Energy Physics, Venice, Italy, July 5–11, 2017.

\bibitem{reno-Neutrino14}
S.-H. Seo, 2014.
\newblock talk given on behalf of the RENO Collaboration at the XXVI
  International Conference on Neutrino Physics and Astrophysics, Boston, USA,
  June 2–7, 2014.

\bibitem{An:2016ses}
{\bf Daya Bay} {\bf Collaboration}, F.~P. An {\em et~al.}, {\it {Measurement of
  Electron Antineutrino Oscillation Based on 1230 Days of Operation of the
  Daya Bay Experiment}},  {\em Phys. Rev.} {\bf D95} (2017), no.~7 072006,
  [\href{http://www.arxiv.org/abs/1610.04802}{{\tt 1610.04802}}].

\bibitem{Gando:2010aa}
{\bf KamLAND} {\bf Collaboration}, A.~Gando {\em et~al.}, {\it {Constraints on
  $\theta_{13}$ from A Three-Flavor Oscillation Analysis of Reactor
  Antineutrinos at KamLAND}},  {\em Phys.Rev.} {\bf D83} (2011) 052002,
  [\href{http://www.arxiv.org/abs/1009.4771}{{\tt 1009.4771}}].

\bibitem{Cleveland:1998nv}
B.~T. Cleveland {\em et~al.}, {\it {Measurement of the solar electron neutrino
  flux with the Homestake chlorine detector}},  {\em Astrophys. J.} {\bf 496}
  (1998) 505--526.

\bibitem{Kaether:2010ag}
F.~Kaether, W.~Hampel, G.~Heusser, J.~Kiko, and T.~Kirsten, {\it {Reanalysis of
  the GALLEX solar neutrino flux and source experiments}},  {\em Phys.Lett.}
  {\bf B685} (2010) 47--54, [\href{http://www.arxiv.org/abs/1001.2731}{{\tt
  1001.2731}}].

\bibitem{Abdurashitov:2009tn}
{\bf SAGE} {\bf Collaboration}, J.~N. Abdurashitov {\em et~al.}, {\it
  {Measurement of the solar neutrino capture rate with gallium metal. III:
  Results for the 2002--2007 data-taking period}},  {\em Phys. Rev.} {\bf C80}
  (2009) 015807, [\href{http://www.arxiv.org/abs/0901.2200}{{\tt 0901.2200}}].

\bibitem{Hosaka:2005um}
{\bf Super-Kamkiokande} {\bf Collaboration}, J.~Hosaka {\em et~al.}, {\it
  {Solar neutrino measurements in Super-Kamiokande-I}},  {\em Phys. Rev.} {\bf
  D73} (2006) 112001, [\href{http://www.arxiv.org/abs/hep-ex/0508053}{{\tt
  hep-ex/0508053}}].

\bibitem{Cravens:2008aa}
{\bf Super-Kamiokande} {\bf Collaboration}, J.~P. Cravens {\em et~al.}, {\it
  {Solar neutrino measurements in Super-Kamiokande-II}},  {\em Phys. Rev.} {\bf
  D78} (2008) 032002, [\href{http://www.arxiv.org/abs/0803.4312}{{\tt
  0803.4312}}].

\bibitem{Abe:2010hy}
{\bf Super-Kamiokande} {\bf Collaboration}, K.~Abe {\em et~al.}, {\it {Solar
  neutrino results in Super-Kamiokande-III}},  {\em Phys.Rev.} {\bf D83} (2011)
  052010, [\href{http://www.arxiv.org/abs/1010.0118}{{\tt 1010.0118}}].

\bibitem{Aharmim:2007nv}
{\bf SNO} {\bf Collaboration}, B.~Aharmim {\em et~al.}, {\it {Measurement of
  the nu/e and total B-8 solar neutrino fluxes with the Sudbury Neutrino
  Observatory phase I data set}},  {\em Phys. Rev.} {\bf C75} (2007) 045502,
  [\href{http://www.arxiv.org/abs/nucl-ex/0610020}{{\tt nucl-ex/0610020}}].

\bibitem{Aharmim:2005gt}
{\bf SNO} {\bf Collaboration}, B.~Aharmim {\em et~al.}, {\it {Electron energy
  spectra, fluxes, and day-night asymmetries of B-8 solar neutrinos from the
  391-day salt phase SNO data set}},  {\em Phys. Rev.} {\bf C72} (2005) 055502,
  [\href{http://www.arxiv.org/abs/nucl-ex/0502021}{{\tt nucl-ex/0502021}}].

\bibitem{Aharmim:2008kc}
{\bf SNO} {\bf Collaboration}, B.~Aharmim {\em et~al.}, {\it {An Independent
  Measurement of the Total Active \iso{B}{8} Solar Neutrino Flux Using an Array
  of \iso{He}{3} Proportional Counters at the Sudbury Neutrino Observatory}},
  {\em Phys. Rev. Lett.} {\bf 101} (2008) 111301,
  [\href{http://www.arxiv.org/abs/0806.0989}{{\tt 0806.0989}}].

\bibitem{Bellini:2011rx}
{\bf The Borexino} {\bf Collaboration}, G.~Bellini {\em et~al.}, {\it
  {Precision measurement of the \iso{Be}{7} solar neutrino interaction rate in
  Borexino}},  {\em Phys.Rev.Lett.} {\bf 107} (2011) 141302,
  [\href{http://www.arxiv.org/abs/1104.1816}{{\tt 1104.1816}}].

\bibitem{Bellini:2008mr}
{\bf Borexino} {\bf Collaboration}, G.~Bellini {\em et~al.}, {\it {Measurement
  of the solar \iso{B}{8} neutrino rate with a liquid scintillator target and 3
  MeV energy threshold in the Borexino detector}},  {\em Phys.Rev.} {\bf D82}
  (2010) 033006, [\href{http://www.arxiv.org/abs/0808.2868}{{\tt 0808.2868}}].

\bibitem{Reichenbacher:2005nc}
J.~Reichenbacher, {\em {Final KARMEN results on neutrino oscillations and
  neutrino nucleus interactions in the energy regime of supernovae}}.
\newblock PhD thesis, Karlsruhe U., 2005.

\bibitem{Armbruster:1998uk}
B.~Armbruster, I.~Blair, B.~Bodmann, N.~Booth, G.~Drexlin, {\em et~al.}, {\it
  {KARMEN limits on $\nu_e \to \nu_\tau$ oscillations in two neutrino and three
  neutrino mixing schemes}},  {\em Phys.Rev.} {\bf C57} (1998) 3414--3424,
  [\href{http://www.arxiv.org/abs/hep-ex/9801007}{{\tt hep-ex/9801007}}].

\bibitem{Conrad:2011ce}
J.~Conrad and M.~Shaevitz, {\it {Limits on Electron Neutrino Disappearance from
  the KARMEN and LSND $\nu_e$--Carbon Cross Section Data}},  {\em Phys.Rev.}
  {\bf D85} (2012) 013017, [\href{http://www.arxiv.org/abs/1106.5552}{{\tt
  1106.5552}}]. Published version.

\bibitem{Auerbach:2001hz}
{\bf LSND} {\bf Collaboration}, L.~Auerbach {\em et~al.}, {\it {Measurements of
  charged current reactions of $\nu_e$ on \iso{C}{12}}},  {\em Phys.Rev.} {\bf
  C64} (2001) 065501, [\href{http://www.arxiv.org/abs/hep-ex/0105068}{{\tt
  hep-ex/0105068}}].

\bibitem{Hampel:1997fc}
{\bf GALLEX} {\bf Collaboration}, W.~Hampel {\em et~al.}, {\it {Final results
  of the Cr-51 neutrino source experiments in GALLEX}},  {\em Phys.Lett.} {\bf
  B420} (1998) 114--126.

\bibitem{Abdurashitov:1998ne}
{\bf SAGE} {\bf Collaboration}, J.~Abdurashitov {\em et~al.}, {\it {Measurement
  of the response of the Russian-American gallium experiment to neutrinos from
  a Cr-51 source}},  {\em Phys.Rev.} {\bf C59} (1999) 2246--2263,
  [\href{http://www.arxiv.org/abs/hep-ph/9803418}{{\tt hep-ph/9803418}}].

\bibitem{Abdurashitov:2005tb}
J.~Abdurashitov, V.~Gavrin, S.~Girin, V.~Gorbachev, P.~Gurkina, {\em et~al.},
  {\it {Measurement of the response of a Ga solar neutrino experiment to
  neutrinos from an Ar-37 source}},  {\em Phys.Rev.} {\bf C73} (2006) 045805,
  [\href{http://www.arxiv.org/abs/nucl-ex/0512041}{{\tt nucl-ex/0512041}}].

\bibitem{Gariazzo:2018mwd}
S.~Gariazzo, C.~Giunti, M.~Laveder, and Y.~F. Li, {\it {Model-Independent
  $\bar\nu_e$ Short-Baseline Oscillations from Reactor Spectral Ratios}},
  \href{http://www.arxiv.org/abs/1801.06467}{{\tt 1801.06467}}.

\bibitem{Maltoni:2003cu}
M.~Maltoni and T.~Schwetz, {\it {Testing the statistical compatibility of
  independent data sets}},  {\em Phys. Rev.} {\bf D68} (2003) 033020,
  [\href{http://www.arxiv.org/abs/hep-ph/0304176}{{\tt hep-ph/0304176}}].

\bibitem{Maltoni:2007zf}
M.~Maltoni and T.~Schwetz, {\it {Sterile neutrino oscillations after first
  MiniBooNE results}},  {\em Phys. Rev.} {\bf D76} (2007) 093005,
  [\href{http://www.arxiv.org/abs/0705.0107}{{\tt 0705.0107}}].

\bibitem{Armbruster:2002mp}
{\bf KARMEN} {\bf Collaboration}, B.~Armbruster {\em et~al.}, {\it {Upper
  limits for neutrino oscillations muon-antineutrino to electron-antineutrino
  from muon decay at rest}},  {\em Phys. Rev.} {\bf D65} (2002) 112001,
  [\href{http://www.arxiv.org/abs/hep-ex/0203021}{{\tt hep-ex/0203021}}].

\bibitem{Astier:2003gs}
{\bf NOMAD} {\bf Collaboration}, P.~Astier {\em et~al.}, {\it {Search for
  $\nu_\mu \to \nu_e$ oscillations in the NOMAD experiment}},  {\em Phys.Lett.}
  {\bf B570} (2003) 19--31,
  [\href{http://www.arxiv.org/abs/hep-ex/0306037}{{\tt hep-ex/0306037}}].

\bibitem{Borodovsky:1992pn}
L.~Borodovsky, C.~Chi, Y.~Ho, N.~Kondakis, W.-Y. Lee, {\em et~al.}, {\it
  {Search for muon-neutrino oscillations $\nu_\mu \to \nu_e$ ($\bar\nu_\mu \to
  \bar\nu_e$) in a wide band neutrino beam}},  {\em Phys.Rev.Lett.} {\bf 68}
  (1992) 274--277.

\bibitem{Antonello:2012pq}
M.~Antonello {\em et~al.}, {\it {Experimental Search for the “LSND Anomaly”
  with the Icarus Detector in the CNGS Neutrino Beam}},  {\em Eur. Phys. J.}
  {\bf C73} (2013), no.~3 2345, [\href{http://www.arxiv.org/abs/1209.0122}{{\tt
  1209.0122}}].

\bibitem{Farnese:2015kfa}
C.~Farnese, {\it {Some recent results from ICARUS}},  {\em AIP Conf. Proc.}
  {\bf 1666} (2015) 110002. see also slides available from
  \url{https://indico.fnal.gov/materialDisplay.py?contribId=265&sessionId=18&materialId=slides&confId=8022}.

\bibitem{MBdataReleaseAPP}
{\bf MiniBooNE} {\bf Collaboration}, ``Data release for arxiv:1207.4809.''
  \url{http://www-boone.fnal.gov/for_physicists/data_release/nue_nuebar_2012/combined.html#fit200}.

\bibitem{Maltoni:2002xd}
M.~Maltoni, T.~Schwetz, M.~A. Tortola, and J.~W.~F. Valle, {\it {Ruling Out
  Four Neutrino Oscillation Interpretations of the LSND Anomaly?}},  {\em Nucl.
  Phys.} {\bf B643} (2002) 321--338,
  [\href{http://www.arxiv.org/abs/hep-ph/0207157}{{\tt hep-ph/0207157}}].

\bibitem{Dydak:1983zq}
F.~Dydak, G.~Feldman, C.~Guyot, J.~Merlo, H.~Meyer, {\em et~al.}, {\it {A
  Search for Muon-neutrino Oscillations in the Delta m**2 Range 0.3-eV**2 to
  90-eV**2}},  {\em Phys.Lett.} {\bf B134} (1984) 281.

\bibitem{AguilarArevalo:2009yj}
{\bf MiniBooNE} {\bf Collaboration}, A.~A. Aguilar-Arevalo {\em et~al.}, {\it
  {A Search for muon neutrino and antineutrino disappearance in MiniBooNE}},
  {\em Phys.Rev.Lett.} {\bf 103} (2009) 061802,
  [\href{http://www.arxiv.org/abs/0903.2465}{{\tt 0903.2465}}].

\bibitem{Cheng:2012yy}
{\bf MiniBooNE Collaboration, SciBooNE Collaboration} {\bf Collaboration},
  G.~Cheng {\em et~al.}, {\it {Dual baseline search for muon antineutrino
  disappearance at $0.1 {\rm eV}^2 < {\Delta}m^2 < 100 {\rm eV}^2$}},  {\em
  Phys.Rev.} {\bf D86} (2012) 052009,
  [\href{http://www.arxiv.org/abs/1208.0322}{{\tt 1208.0322}}].

\bibitem{Wendell:2010md}
{\bf Super-Kamiokande Collaboration} {\bf Collaboration}, R.~Wendell {\em
  et~al.}, {\it {Atmospheric neutrino oscillation analysis with sub-leading
  effects in Super-Kamiokande I, II, and III}},  {\em Phys.Rev.} {\bf D81}
  (2010) 092004, [\href{http://www.arxiv.org/abs/1002.3471}{{\tt 1002.3471}}].

\bibitem{MINOS:2016viw}
{\bf MINOS} {\bf Collaboration}, P.~Adamson {\em et~al.}, {\it {Search for
  Sterile Neutrinos Mixing with Muon Neutrinos in Minos}},  {\em Phys. Rev.
  Lett.} {\bf 117} (2016), no.~15 151803,
  [\href{http://www.arxiv.org/abs/1607.01176}{{\tt 1607.01176}}].

\bibitem{Dziewonski:1981xy}
A.~M. Dziewonski and D.~L. Anderson, {\it {Preliminary Reference Earth Model}},
   {\em Phys. Earth Planet. Interiors} {\bf 25} (1981) 297--356.

\bibitem{MBdataReleaseDIS}
{\bf MiniBooNE} {\bf Collaboration}, ``Data release for arxiv:0903.2465.''
  \url{http://www-boone.fnal.gov/for_physicists/data_release/numu_numubar/}.

\bibitem{Blennow:2018hto}
M.~Blennow, E.~Fernandez-Martinez, J.~Gehrlein, J.~Hernandez-Garcia, and
  J.~Salvado, {\it {IceCube bounds on sterile neutrinos above 10 eV}},
  \href{http://www.arxiv.org/abs/1803.02362}{{\tt 1803.02362}}.

\bibitem{Bilenky:1996rw}
S.~M. Bilenky, C.~Giunti, and W.~Grimus, {\it {Neutrino Mass Spectrum from the
  Results of Neutrino Oscillation Experiments}},  {\em Eur. Phys. J.} {\bf C1}
  (1998) 247--253, [\href{http://www.arxiv.org/abs/hep-ph/9607372}{{\tt
  hep-ph/9607372}}].

\bibitem{Okada:1996kw}
N.~Okada and O.~Yasuda, {\it {A Sterile Neutrino Scenario Constrained by
  Experiments and Cosmology}},  {\em Int. J. Mod. Phys.} {\bf A12} (1997)
  3669--3694, [\href{http://www.arxiv.org/abs/hep-ph/9606411}{{\tt
  hep-ph/9606411}}].

\bibitem{Barger:1998bn}
V.~D. Barger, S.~Pakvasa, T.~J. Weiler, and K.~Whisnant, {\it {Variations on
  Four Neutrino Oscillations}},  {\em Phys. Rev.} {\bf D58} (1998) 093016,
  [\href{http://www.arxiv.org/abs/hep-ph/9806328}{{\tt hep-ph/9806328}}].

\bibitem{Cyburt:2015mya}
R.~H. Cyburt, B.~D. Fields, K.~A. Olive, and T.-H. Yeh, {\it {Big Bang
  Nucleosynthesis: 2015}},  {\em Rev. Mod. Phys.} {\bf 88} (2016) 015004,
  [\href{http://www.arxiv.org/abs/1505.01076}{{\tt 1505.01076}}].

\bibitem{Ade:2015xua}
{\bf Planck} {\bf Collaboration}, P.~A.~R. Ade {\em et~al.}, {\it {Planck 2015
  results. XIII. Cosmological parameters}},  {\em Astron. Astrophys.} {\bf 594}
  (2016) A13, [\href{http://www.arxiv.org/abs/1502.01589}{{\tt 1502.01589}}].

\bibitem{Dasgupta:2013zpn}
B.~Dasgupta and J.~Kopp, {\it {A m\'enage \`a trois of eV-scale sterile
  neutrinos, cosmology, and structure formation}},  {\em Phys.Rev.Lett.} {\bf
  112} (2014) 031803, [\href{http://www.arxiv.org/abs/1310.6337}{{\tt
  1310.6337}}].

\bibitem{Hannestad:2013ana}
S.~Hannestad, R.~S. Hansen, and T.~Tram, {\it {How secret interactions can
  reconcile sterile neutrinos with cosmology}},  {\em Phys.Rev.Lett.} {\bf 112}
  (2014) 031802, [\href{http://www.arxiv.org/abs/1310.5926}{{\tt 1310.5926}}].

\bibitem{Chu:inprogress}
X.~Chu, B.~Dasgupta, M.~Dentler, J.~Kopp, and N.~Saviano.
\newblock work in progress.

\bibitem{Cherry:2016jol}
J.~F. Cherry, A.~Friedland, and I.~M. Shoemaker, {\it {Short-baseline neutrino
  oscillations, Planck, and IceCube}},
  \href{http://www.arxiv.org/abs/1605.06506}{{\tt 1605.06506}}.

\bibitem{Yaguna:2007wi}
C.~E. Yaguna, {\it {Sterile neutrino production in models with low reheating
  temperatures}},  {\em JHEP} {\bf 06} (2007) 002,
  [\href{http://www.arxiv.org/abs/0706.0178}{{\tt 0706.0178}}].

\bibitem{Saviano:2013ktj}
N.~Saviano, A.~Mirizzi, O.~Pisanti, P.~D. Serpico, G.~Mangano, {\em et~al.},
  {\it {Multi-momentum and multi-flavour active-sterile neutrino oscillations
  in the early universe: role of neutrino asymmetries and effects on
  nucleosynthesis}},  {\em Phys.Rev.} {\bf D87} (2013) 073006,
  [\href{http://www.arxiv.org/abs/1302.1200}{{\tt 1302.1200}}].

\bibitem{Ho:2012br}
C.~M. Ho and R.~J. Scherrer, {\it {Sterile Neutrinos and Light Dark Matter Save
  Each Other}},  {\em Phys.Rev.} {\bf D87} (2013) 065016,
  [\href{http://www.arxiv.org/abs/1212.1689}{{\tt 1212.1689}}].

\bibitem{Giovannini:2002qw}
M.~Giovannini, H.~Kurki-Suonio, and E.~Sihvola, {\it {Big bang nucleosynthesis,
  matter antimatter regions, extra relativistic species, and relic
  gravitational waves}},  {\em Phys. Rev.} {\bf D66} (2002) 043504,
  [\href{http://www.arxiv.org/abs/astro-ph/0203430}{{\tt astro-ph/0203430}}].

\bibitem{Bernal:2016gxb}
J.~L. Bernal, L.~Verde, and A.~G. Riess, {\it {The trouble with $H_0$}},  {\em
  JCAP} {\bf 1610} (2016), no.~10 019,
  [\href{http://www.arxiv.org/abs/1607.05617}{{\tt 1607.05617}}].

\bibitem{Lhuillier:Moriond2018}
{\bf STEREO} {\bf Collaboration}, D.~Lhuillier, ``Nuclear physics and the
  reactor anomaly.''
\newblock talk given on behalf of the STEREO collaboration at the 53rd
  Rencontres de Moriond, Mar 10--17, 2018. slides availbale at
  \url{https://indico.in2p3.fr/event/16579/contributions/60857/attachments/47355/59539/01_Lhuillier_Moriond2018_.pdf}.

\bibitem{Huber:2004ka}
P.~Huber, M.~Lindner, and W.~Winter, {\it {Simulation of Long-Baseline Neutrino
  Oscillation Experiments with Globes (General Long Baseline Experiment
  Simulator)}},  {\em Comput. Phys. Commun.} {\bf 167} (2005) 195,
  [\href{http://www.arxiv.org/abs/hep-ph/0407333}{{\tt hep-ph/0407333}}].

\bibitem{Huber:2007ji}
P.~Huber, J.~Kopp, M.~Lindner, M.~Rolinec, and W.~Winter, {\it {New Features in
  the Simulation of Neutrino Oscillation Experiments with Globes 3.0: General
  Long Baseline Experiment Simulator}},  {\em Comput. Phys. Commun.} {\bf 177}
  (2007) 432--438, [\href{http://www.arxiv.org/abs/hep-ph/0701187}{{\tt
  hep-ph/0701187}}].

\end{thebibliography}\endgroup

\end{document}